\documentclass[pre,twocolumn,showpacs,superscriptaddress]{revtex4}
\usepackage{amsmath,amssymb}
\usepackage[pdftex]{graphicx}

\begin{document}
\title{
Comparative
quantum and semi-classical analysis  of Atom-Field Systems II: chaos and regularity}

\author{M. A. Bastarrachea-Magnani}
\affiliation{Instituto de Ciencias Nucleares,
Universidad Nacional Aut\'onoma de M\'exico, Apdo. Postal 70-543, Mexico D. F., C.P. 04510}
\author{S. Lerma-Hern\'andez}
\affiliation{Departamento de F\'isica, Universidad Veracruzana, Circuito Aguirre Beltr\'an s/n, Xalapa, Veracruz, M\'exico, C.P. 91000}
\email{slerma@uv.mx}
\author{J. G. Hirsch}
\affiliation{Instituto de Ciencias Nucleares,
Universidad Nacional Aut\'onoma de M\'exico, Apdo. Postal 70-543, Mexico D. F., C.P. 04510} 

\date{}
\begin{abstract}
The  non-integrable Dicke model and its integrable approximation, the Tavis-Cummings (TC) model, are studied 
as functions of both the coupling constant and the excitation energy. 
The present contribution extends the 
analysis presented
in the previous paper
by focusing
on the statistical properties of the quantum fluctuations in the energy spectrum and their relation with the excited state quantum phase transitions (ESQPT).
These properties are compared with the dynamics observed in the semi-classical versions of the models.   The presence of chaos for different energies and coupling constants is exhibited,  employing Poincar\'e sections and Peres lattices in the classical and  quantum versions, respectively. A clear correspondence between the classical and quantum result is found for systems 
containing between $\mathcal{N} = 80$ to $200$ atoms.
A  measure of the Wigner character of the energy  spectrum 
for
different couplings and energy intervals  is also presented 
employing
the statistical  Anderson-Darling test.  It is found that  in the Dicke Model, for any coupling,  a low energy regime with regular states is always present.  The richness of the onset of chaos is discussed both for finite quantum systems and for the semi-classical limit, 
which is exact when the number of atoms in the system tends to infinite.
\end{abstract}
\maketitle


\noindent


\section{Introduction}

The Dicke and Tavis-Cummings (TC)  Hamiltonian describe a system of $\mathcal{N}$ two-level atoms interacting with a single monochromatic electromagnetic radiation mode within a cavity \cite{Dicke54}. One of  the most representative feature of these  Hamiltonians is their second-order quantum phase transition (QPT) in the thermodynamic limit \cite{Hepp73,Wang73} (equivalent in the present models to the semi-classical limit). The ground state of the system goes from a normal to a superradiant state when the atom-field interaction reaches a critical value. In a 
companion
paper \cite{Basta1}, hereon referred to as (I), it was shown that the semi-classical approximation to the density of states (DoS) describes very well the averaged quantum density of states (QDoS). From the semi-classical 
description,
the presence of two different  excited-state quantum phase transitions (ESQPTs) was clearly established in these models. One ESQPT, referred to as {\em static}, occurs for any coupling at  an energy $E/(\omega_o j)=1$,  where  the whole phase space associated with the two-levels atoms (the pseudo-spin sphere)  becomes available for the system.  The second ESQPT, referred to as {\em dynamic}, 
can take place
only in the superradiant phase, at energies $E/(\omega_o j)=-1$. This transition occurs when the top of the double well (Dicke) or mexican hat (TC)  potential  that develops in the superradiant phase is attained. 

The previous results allow to study the properties of the quantum fluctuations using the semi-classical DoS to separate the tendency or secular variation of the spectrum from its fluctuations. It is known that the tendency of the QDoS depends completely on the particular system, but the properties of the fluctuations are universal \cite{Haake}. For quantum systems with a classical analogue integrable the fluctuations are common with those of the so-called Gaussian Diagonal Ensemble (GDE), while for time-reversal-symmetric quantum systems with classical analogue hard chaotic, the fluctuations are those of the Gaussian Orthogonal Ensemble (GOE). 

In Ref. \cite{Emary03} numerical evidence 
was presented 
that  suggests a relationship  between the normal to superradiant  phase transition  and the onset of chaos in the non-integrable Dicke model. More recently \cite{Per11A} this relationship  was studied more closely and it was suggested  that the onset of chaos is caused by the precursors of the {\em dynamic} ESQPT occurring  in the superradiant phase. In this contribution we go further in the study of this relationship  between the onset of chaos and singular  behavior of the DoS (ESQPT) and ground-state energy (QPT).
To this end we consider the non-integrable Dicke model and its integrable approximation, the Tavis-Cummings model.  We study these models as a function of both the coupling between atoms and field, and the
excitation
 energy.  The presence of chaos for different energies and coupling constants is exhibited, in the semi-classical limit, employing Poincar\'e sections. The role of the classical chaos in the Dicke model has been recently  studied  in the context of the equilibration of unitary quantum dynamics \cite{Alt12}. In the quantum case  Peres lattices are presented, which allow to characterize regular and chaotic regions qualitatively. 
A quantitative measure of the properties of the energy spectrum is also presented by means  of testing if the Nearest Neighbor Spacing Distribution (NNSD) of the unfolded energies follows the Wigner distribution of the GOE. The results of the classical model are compared to the quantum ones. Similar to the results of Ref.\cite{Bakemeier13}, a clear correspondence between classical and quantum results are found for finite quantum system ranging from $\mathcal{N}\sim 80$ to $\mathcal{N}\sim 200$. 

 It is found that the onset of chaos in the Dicke model 
can  only  take place in the energy region where  
 the quadratic approximation of the Hamiltonian,
 the one
 obtained by considering small oscillations around the global energy minimum,
 fails to describe semiclassical model dynamics.
 For any coupling  there always exist 
 a low
 energy interval above the 
 ground state
 where only regular patterns are 
 observed.
In particular, for the very small coupling regime $\gamma\approx  0$ this energy interval extends to infinity. Above these low energy 
region
the quadratic approximation breaks 
down
and room is left for the onset of chaos. Even if an indirect connection between the ESQPTs and the onset of chaos can be identified through the unstable fixed points, the onset of chaos is a much richer phenomenon than the occurrence of non-analytic behavior in the density of states or the ground-state energy.
 
The article is organized as follows: in Section II we present briefly the Quantum Dicke and Tavis-Cummings Hamiltonians and their classical analogues, and summarize some of their properties. In section III  a qualitatively analysis of the spectrum is done via Peres lattices and Poincar\'e sections,  revealing a clear classical and quantum correspondence 
at
the onset of chaos, both in the normal and superradiant phases.
The properties of the energy fluctuations  
are studied
by means of the Anderson-Darling test for the NNSD against the Wigner distribution.  Section IV contains the conclusions. 

\section{Dicke and Tavis-Cummings Hamiltonians}

The Dicke and Tavis-Cummings Hamiltonians are made of three parts: one associated to the monochromatic quantized radiation field (boson operators $a$ and $a^\dagger$), a second one to the atomic sector (pseudo-spin operators $J_z$ and $J_\pm$), and a last one which describes the interaction between them 
\begin{equation}
\begin{split}
H&=\omega a^{\dagger}a+\omega_{0}J_{z}+\\
&+\frac{\gamma}{\sqrt{\mathcal{N}}}\left[ \left(aJ_{+}+a^{\dagger}J_{-}\right) +\delta\left(a^\dagger J_{+}+a J_{-}\right)\right],
\end{split}
\end{equation}
where $\delta=0$ and $1$ for the TC and Dicke models, respectively. A quantum phase transition, from the normal to the so-called superradiant phase, takes place at a value of the coupling constant given by  $\gamma_{c}=\sqrt{\omega_{0}\omega}/(1+\delta)$.  We will focus on the subspace with largest pseudo-spin, where $j=\mathcal{N}/2$ \cite{Nah13}. 

The TC Hamiltonian is integrable because it commutes with the $\Lambda$ operator, $\Lambda=a^{\dagger}a+J_{z}+j$. Its conserved eigenvalues $\lambda$ define a set of subspaces where the TC Hamiltonian can be diagonalized independently. The Dicke Hamiltonian is not integrable, but  its Hilbert space can be separated  in two sectors depending on the eigenvalue ($p=\pm$) of the Parity operator $\Pi=e^{i\pi\Lambda}$.

The classical version of the Dicke and TC  models can be obtained employing the {\it naive} substitution  \cite{DickeBrasil}  of the pseudospin variables by classical angular momentum ones ($J_i\rightarrow j_i$), and the substitution of the boson variables by a classical harmonic oscillator (variables $q$ and $p$) with $m\omega=1$. The pseudospin  variables satisfy  the  Poisson-bracket algebra $\{j_i,j_j\}=\epsilon_{ijk} j_k$. From there canonical variables $\{P,Q\}=-1$ can be obtained as $P=j_z$ and $Q=\phi=\tan^{-1}(j_y/j_x)$, where $\phi$ is the azimuthal angle of the vector $\vec j=(j_x,j_y,j_z)$ whose magnitude is constant $|\vec j|=j$. In terms of the canonical variables the classical Dicke Hamiltonian reads 
\begin{eqnarray}
H_{cl}&=& \omega_o j_z+\frac{\omega}{2}(q^2+p^2)+ \label{DiHam}\\
      &  &\gamma \sqrt{j} \sqrt{1-\frac{j_z^2}{j^2}}\left[ (1+\delta) q \cos\phi -(1-\delta) p\sin\phi\right].\nonumber
\end{eqnarray} 
From here the equations of motion are 
\begin{eqnarray}
\frac{dq}{dt}&=&\frac{\partial H_{cl}}{\partial p}= \omega p- (1-\delta)\gamma \sqrt{j} \sqrt{1-\frac{j_z^2}{j^2}}  \sin\phi \label{qp}\\
\frac{dp}{dt}&=&-\frac{\partial H_{cl}}{\partial q}=-q\omega -(1+\delta)\gamma\sqrt{j} \sqrt{1-\frac{j_z^2}{j^2}}\cos\phi  \label{pp}\\
\frac{d\phi}{dt}&=&\frac{\partial H_{cl}}{\partial j_z}=\omega_o  \\
& -&\frac{\gamma j_z }{j^{3/2} \sqrt{1-\frac{j_z^2}{j^2}}} \left[ (1+\delta) q \cos\phi -(1-\delta) p\sin\phi\right] \label{php}\nonumber\\
\frac{d j_z}{dt}&=&-\frac{\partial H_{cl}}{\partial \phi}=2\gamma\sqrt{j}\sqrt{1-\frac{j_z^2}{j^2}}\nonumber\\
 &\times& \left[(1+\delta) q\sin\phi +(1-\delta) p\cos\phi\right].\label{jp}
\end{eqnarray} 
\begin{figure*}
\begin{tabular}{cc}
\rotatebox{90}{\qquad \qquad\qquad  \Large{$E/(\omega_o j)$}}\includegraphics[width=0.4 \textwidth]{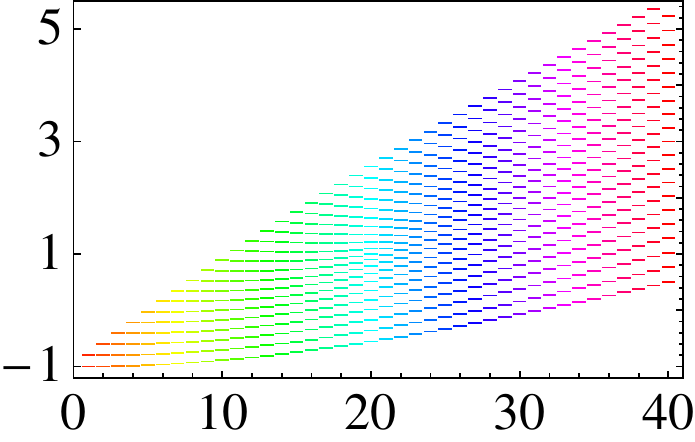}&
\rotatebox{90}{\qquad \qquad\qquad  \Large{$\Delta E$}}\includegraphics[width=0.4 \textwidth]{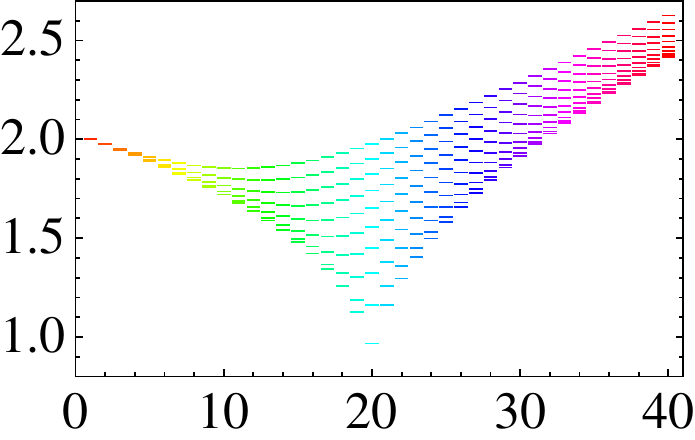}
\\
\rotatebox{90}{\qquad \qquad\qquad  \Large{$E/(\omega_o j)$}}\includegraphics[width=0.4 \textwidth]{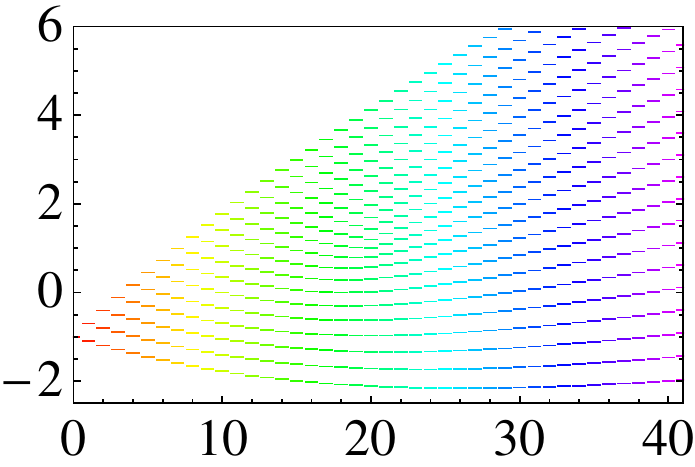}&
\quad\rotatebox{90}{\qquad \qquad\qquad  \Large{$\Delta E$}}\quad\includegraphics[width=0.385 \textwidth]{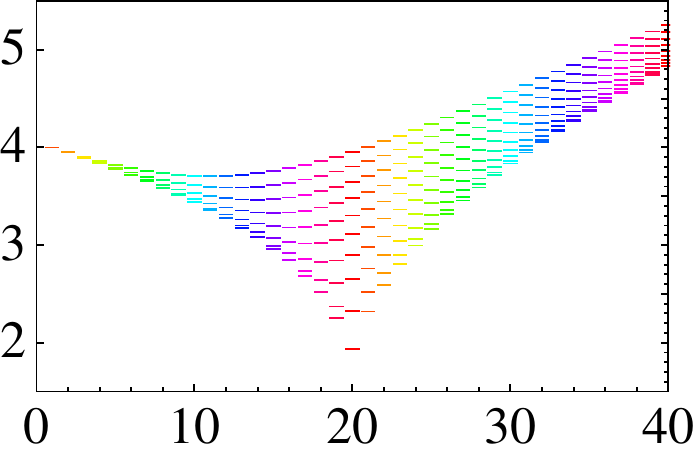}\\
\qquad\quad\Large{$\lambda$}&\qquad\quad\Large{$\lambda$}
\end{tabular}
\caption{(Color online) Energy (left) and energy difference (right) as functions of $\lambda$. The parameters are $j=10$, $n_{max}=10$, $\lambda_{max}=60$, $\omega=\omega_{0}=1$, $\gamma= \gamma_{c}$ (above) and $\gamma= 2\gamma_{c}$ (below)}
\label{fig7}
\end{figure*}
Employing
the  semi-classical approximation to the density of states for a given energy, 
$$\nu(E)=\frac{1}{(2 \pi)^2}\int dq \, dp\, d\phi\, d j_z\,  \delta(E-H_{cl}(q,p,\phi,j_z)),$$
it is possible [see (I)] to identify two ESQPTs in the energy space. One, referred to as {\em static}, occurs at an energy $E/(\omega_o j)=1$  for any coupling. The second one, referred to as {\it dynamic}, takes place only in the superradiant phase at energies  $E/(\omega_o j)=-1$. The semiclassical approximation to the density of states in the Dicke model is given by \cite{Basta1,Bran13}
\begin{equation}
\frac{\omega}{2 j} \nu(\epsilon)= \left\{ 
\begin{array}{l}
\frac{1}{\pi}\int_{y_-}^{y_+} \arccos \sqrt{\frac{2\gamma_c^2 (y-\epsilon)}{\gamma^2(1-y^2)}} dy,   \ \ \ \ \epsilon_0 \leq \epsilon < -1 \\
\frac{\epsilon+1}{2}+\frac{1}{\pi}\int_{\epsilon}^{y_+} \arccos \sqrt{\frac{2\gamma_c^2 (y-\epsilon)}{\gamma^2(1-y^2)}} dy  , \ \ \ |\epsilon|\leq 1 \\
 1,\ \ \ \ \ \ \ \ \ \ \ \ \ \ \ \ \ \ \ \ \ \ \ \ \ \  \ \ \ \ \ \ \  \ \ \  \ \ \ \ \ \ \  \ \ 	\epsilon > 1 ,
\end{array}
\right.
\label{dosDicke}
\end{equation} 
where $ 
y_{\pm}= \left(-\frac{\gamma_c^2}{\gamma^2} \pm \frac{\gamma_c}{\gamma}\sqrt{2(\epsilon-\epsilon_o)} \right)$, with $\epsilon_o\equiv-\frac{1}{2}\left(\frac{\gamma_c^ 2}{\gamma^2}+\frac{\gamma^2}{\gamma_c^2}\right)$, and we have defined the scaled energy $\epsilon=E/(\omega_o j)$.
As it is discussed in (I), the singular behavior of the density of states can be related with the unstable points of the Hamiltonian classical flux. The relationship between the ESQPT and  the onset of chaos is discussed in the following section.    

\section{Regularity and chaos}

To establish the onset of chaos in the classical version of the models, we use Poincar\'e surface sections for different couplings and energies. All the Poincar\'e sections shown along this contribution  were obtained  as follows: we solved numerically the equations of motion, Eqs.(\ref{qp}-\ref{jp}), and   considered intersections of the orbits with the surface $p=0$. The intersections define a two dimensional surface in the three-dimensional space $q$-$j_z$-$\phi$. For given ($E,j_z$, $\phi$) the energy conservation, $E=H_{cl}(p=0,q,j_z,\phi)$, gives two possible values $q_{\pm}$ for the variable $q$. We selected points corresponding to the largest $q_{\pm}$ and projected them  finally  in the polar plane $[1+(j_z/j)]$-$\phi$. For the quantum versions we use Peres lattices \cite{Per84}, which are a  visual method that plays a role similar to that of the Poincar\'e sections in classical mechanics. The Peres lattices are very useful to study the route to chaos in quantum systems of two degrees of freedom. If a quantum system with two degrees of freedom  and unperturbed Hamiltonian $H_{0}$ is integrable, a plot between the 
Hamiltonian  eigenenergies 
and the respective eigenvalues of the  constant of motion $I$ ( $[H_{0},I]=0$) form a lattice of regularly distributed points, because each energy level has a natural way to be labeled by the quantum number associated to $I$. When the system is perturbed, $H=H_{0}+\gamma H_{1}$ and becomes non-integrable, $I$ is not longer  a conserved quantity. However, we can use the expectation values of  $I$ (the  Peres operator) in the energy eigenstates  and plot them against the Hamiltonian eigenvalues. This choice connects the unperturbed and perturbed cases and such plots are called Peres lattices. A small perturbation does not destroy the regular lattice of the integrable case, instead a localized distortion is created in the lattice while  the rest of the lattice remains regular. As the perturbation increases the irregular part of the lattice increases as well, allowing to identify in a simple way the regions with classical  chaotic, regular or mixed counterpart. In this way the Peres method represents a qualitatively sensitive probe that allows to visualize  the competition between regular and chaotic behavior in the quantum spectrum of a system \cite{Str09}. Moreover, the freedom in choosing the Peres operator makes it possible to focus on various properties of individual states and to closely follow the way how chaos sets in and proliferates in the system. The Peres lattices  help us not only to characterize the chaotic, regular and mixed regimes in the quantum spectrum, they allow us to qualitatively identify the ESQPT's and their properties. 
 \begin{figure}
\centering
\begin{tabular}{c}
\rotatebox{90}{\qquad \qquad\qquad  \Large{$\langle J_z\rangle/j$}}\includegraphics[width=0.41 \textwidth]{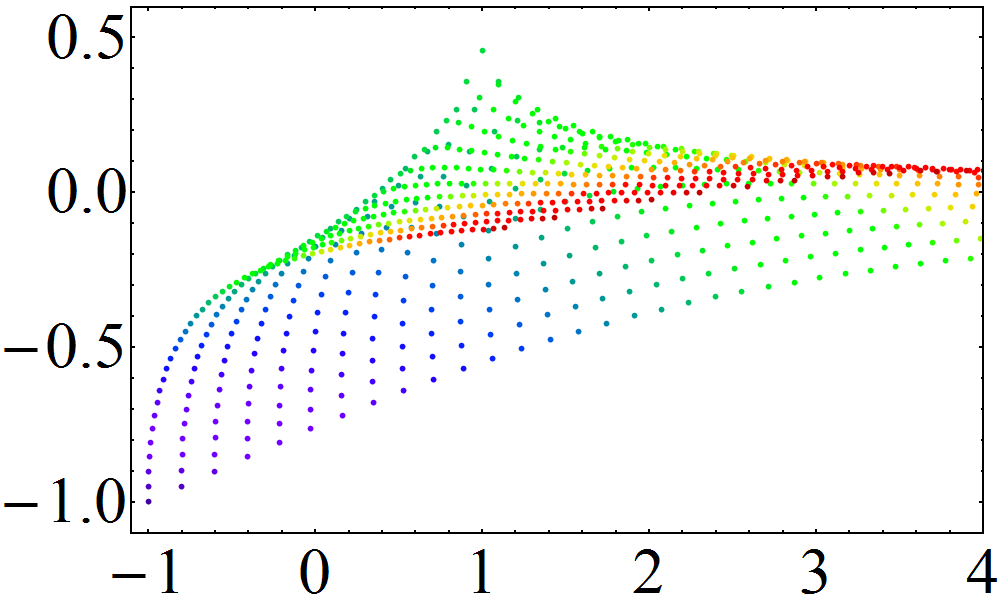}
\\
\rotatebox{90}{\qquad \qquad\qquad  \Large{$\langle J_z\rangle/j$}}\includegraphics[width=0.41 \textwidth]{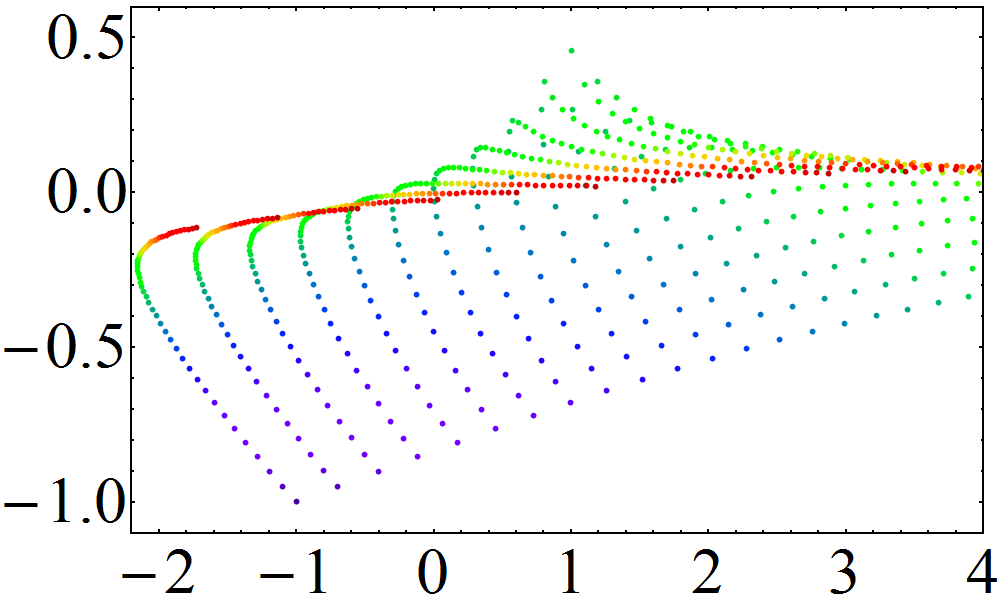}
\\
\qquad\qquad\qquad\Large{$E/(\omega_o j)$}
\end{tabular}
\caption{(Color online) Peres lattices for the Tavis-Cummings model, for $\frac{\langle J_z \rangle}{j} $. The parameters are $j=10$, $\lambda_{max}=50$, $\omega=\omega_{0}=1$,  $\gamma=\gamma_{c}$ (above) and $\gamma=2 \gamma_{c}$ (below). Each color represents states  with the same  $\lambda$.}
\label{fig8}
\end{figure}

\begin{figure}
\begin{tabular}{ll}
\vtop{\vskip -0.15\textwidth \hbox{\includegraphics[width=0.2 \textwidth]{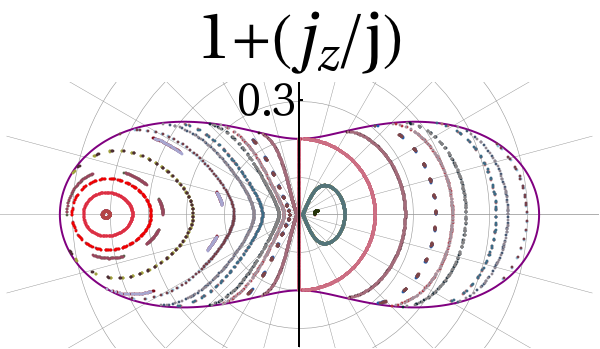}
}}&
\includegraphics[width=0.13 \textwidth]{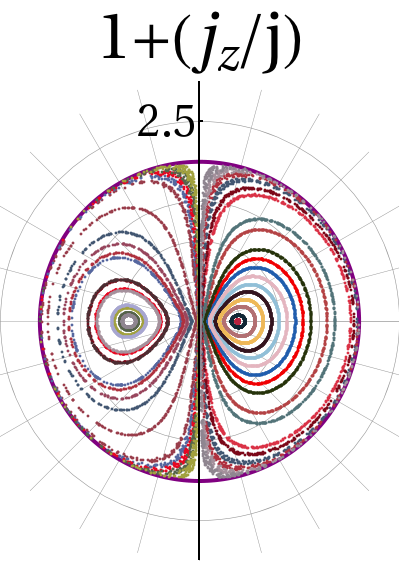}
\\
\vtop{\vskip -0.165\textwidth \hbox{\includegraphics[width=0.2 \textwidth]{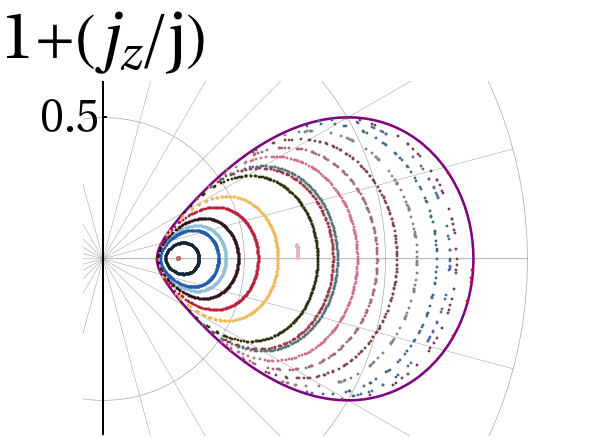}
}}&
\includegraphics[width=0.13 \textwidth]{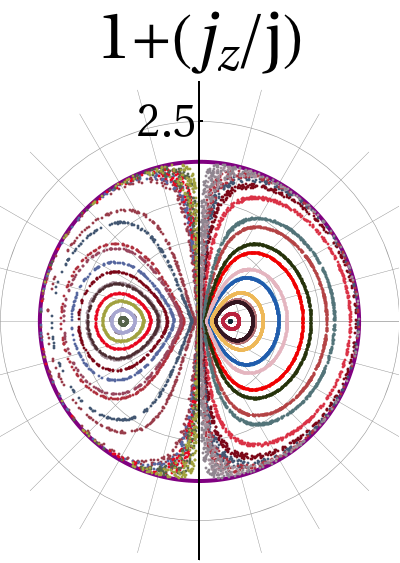}
\end{tabular}
\caption{(Color online) Poincar\'e sections in polar coordinates of the  pseudospin variables  ($\phi$ and $1+j_z/j$) for the classical Tavis-Cummings model for  the same couplings as  fig. \ref{fig8} ($\gamma=\gamma_{c}$ top row and $\gamma=2 \gamma_{c}$ bottom row). The energies used are $E/(\omega_o j)=-0.8$ and $1.2$ for the top row,    and  $E/(\omega_o j)=-1.5$ and $1.5$ for the bottom one.}
\label{fig8b}
\end{figure}
\begin{figure*}
\begin{tabular}{ccc}
\rotatebox{90}{\qquad \qquad  \large{$\langle J_z\rangle/j$}}\includegraphics[width=0.303 \textwidth]{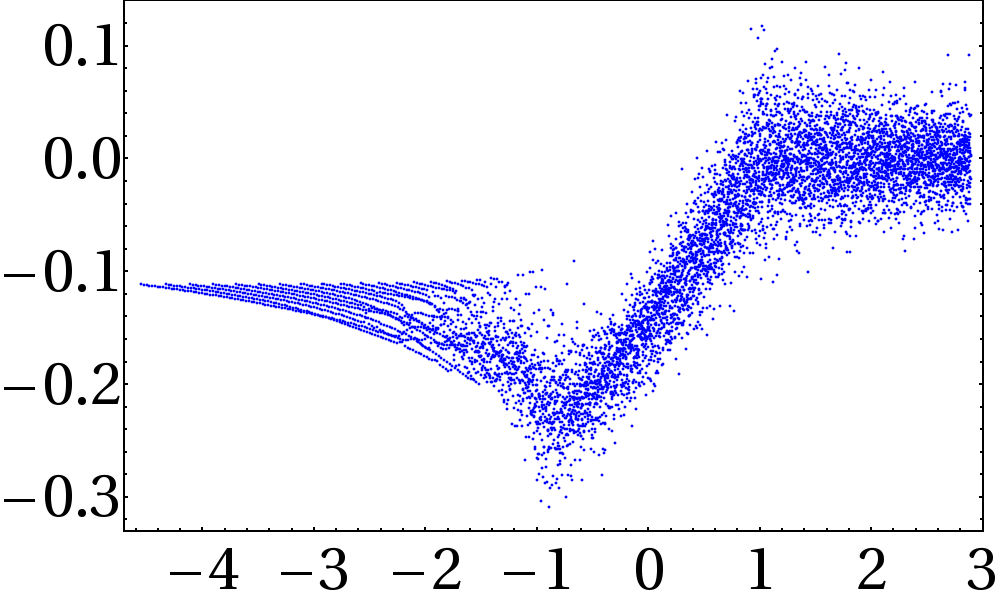}
&\rotatebox{90}{\qquad \qquad  \large{$\langle J_x^2\rangle/j$}}\ \includegraphics[width=0.287 \textwidth]{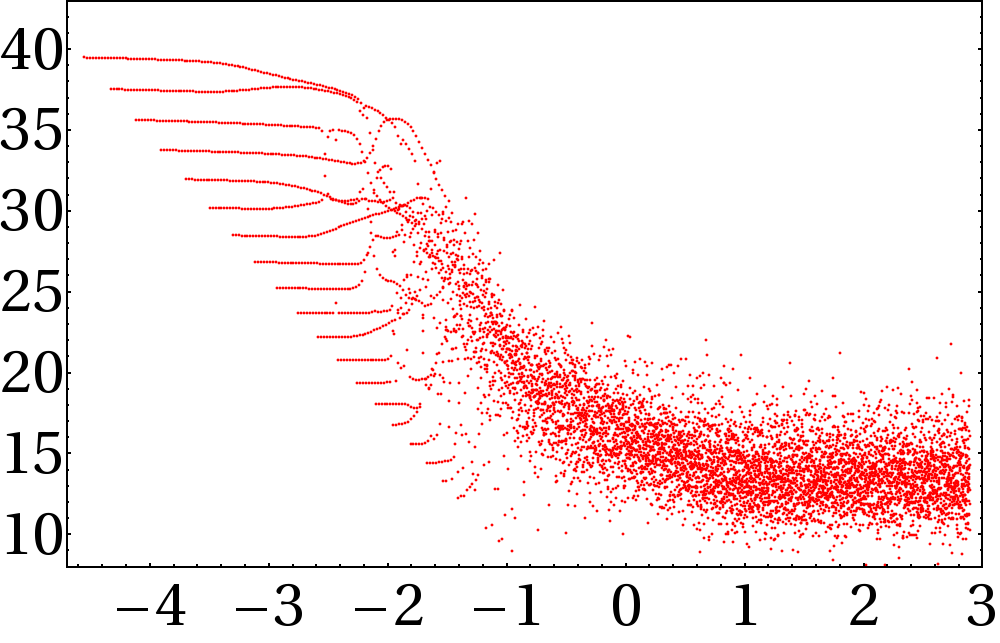}
&\rotatebox{90}{\qquad \qquad  \large{$\langle a^\dagger a \rangle/j$}}\ \includegraphics[width=0.300 \textwidth]{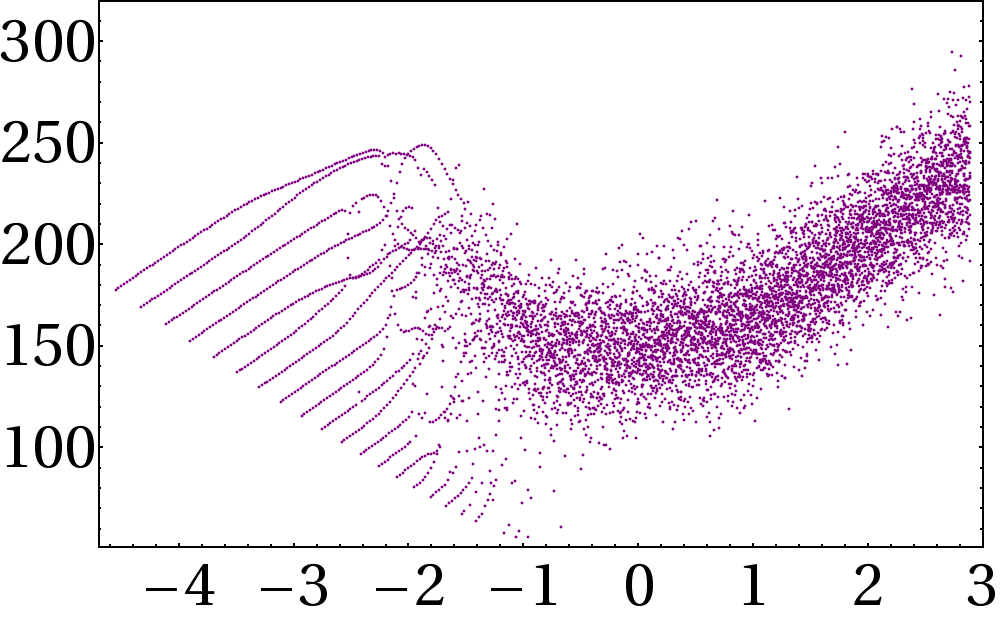}
\\
\Large{$E/(\omega_o j)$} & \Large{$E/(\omega_o j)$}& \Large{$E/(\omega_o j)$}
\end{tabular}
\caption{(Color online) Peres lattices of the Dicke model using $J_{z}$(left), $J_{x}^{2}$ (center) and $a^{\dagger}a$ (right) as Peres operators. The parameters are $j=40$, $\omega=\omega_{0}=1$, $\gamma=3\gamma_{c}$ and $N_{max}=300$ [see (I)]. Only positive parity states are considered.}
\label{fig9}
\end{figure*} 
 
  Besides the Peres lattices of the quantum models, we analyze the  statistical properties of  their  energy  spectra,  in different energy intervals and for different couplings. 
To this end
we test if the energy fluctuations follow the Wigner distribution \cite{Haake}, which is typical of the quantum systems with classical counterpart chaotic. 
The characterization is performed employing
the Anderson-Darling test \cite{Anderson}, which gives a simple criterion to establish if a given empirical set of data follows a given  theoretical distribution.   Before presenting results for the non-integrable Dicke model, the Tavis-Cummings model is discussed. In particular we focus  on the signatures of the ESQPT [the abrupt changes in the DoS calculated classically in (I)] appearing in the Peres lattices of the model.

\subsection{Integrable case: Tavis-Cummings}

It is instructive to plot the Peres lattice for the energy $E/(\omega_o j)$  in the Tavis-Cummings model against  the excitation number $\lambda$, displayed in Fig. \ref{fig7} (left) for  $\gamma=\gamma_{c}$ (up) and $\gamma= 2\, \gamma_{c}$ (down), in the resonant case  $\omega=\omega_{o}=1$, for a $j=10$ system. In both plots a {\em dislocation} in the regular lattice can be observed at the point with coordinates  $ E/(\omega_o j) = 1$  and $\lambda=2j$ (20 in this example). It corresponds to the point in which the {\em static} excited state phase transition takes place. On the right hand side of the same figure,  the energy differences between successive levels, for each value of $\lambda$, are shown. It helps to identify the point where these differences have a minimum. This is another way to recognize the presence of a singularity in the DoS, presented in (I).

When we choose $J_{z}$ as Peres operator, we obtain the Peres lattices shown in Fig.\ref{fig8} for two representative values of the coupling, $\gamma=\gamma_c$ and $\gamma=2\gamma_c$ in resonance $\omega=\omega_{0}=1$. It can be seen from the figure that the Peres lattices  are regular (as expected due to  the integrability of the TC model). In the  same figure the  two ({\em static} and {\em dynamic}) ESQPT are clearly 
evidenced 
by  peaks in the lattices at energies $E/(\omega_o j)=-1$ and  $1$. The ESQPT at the energy $E/(\omega_{o}j)=1$, was reported originally by Perez-Fern\'andez et. al. \cite{Per11A} and is associated  to the  state with  $\lambda=2 j$, and maximum expectation value of $\langle J_{z} \rangle$. It corresponds to the   {\em static} ESQPT found in (I), when the energy is equal to that of the unstable fixed point in the north pole of the pseudospin sphere and the  whole pseudospin sphere becomes   accessible. This ESQPT appears for any coupling below or above $\gamma_c$.  The other ESQPT, that takes place at $E/(\omega_{o}j)=-1$, appears only for $\gamma>\gamma_c$ and  is associated with the state with $\lambda = 0$, i.e. with no photons and no excited atoms (the ground state  in the non-interacting case). From a classical point of view it corresponds to the unstable fixed point that develops at the south pole [see companion paper (I)] of the pseudospin sphere when $\gamma>\gamma_c$. The  peaks in the lattice Peres of the TC model  will be also present in the Dicke model, showing that the Peres lattices are able to detect in a visual and simple way, singular behaviors in the models.  Poincar\'e sections for the classical version of the TC are shown in Fig. \ref{fig8b}. As expected for the TC model, the Poincar\'e sections give exclusively regular orbits in accord with the regular patterns of the quantum Peres lattices.

\subsection{The non-integrable Dicke model}

For the Dicke model, where the $\Lambda$ symmetry of the TC model is broken, more complex  Peres lattices are expected. In figure \ref{fig9} we present  Peres lattices for the Dicke model in the superradiant phase ($\gamma=3\gamma_c$),  using $J_{z}$, $J^2_x$ and $ a^\dagger a$ as  Peres operators.  In the three lattices a regular region in the lower part of the spectrum can be clearly identified. For  energies  $(E/\omega_o j)\approx -2$ the regularity begins to disappear and an irregular pattern is established  instead.  For larger energies the lattice is completely irregular. This "route to chaos"  in the quantum results, already  sketched  by Perez-Fern\'andez, et. al. \cite{Per11A},  has a classical correspondence as it is  discussed below.  Before, it is worth mentioning that, as in the case of the TC model, signatures of the {\em static} and {\em dynamic} ESQPTs are clearly seen in the lattice with $J_z$ as Peres operator. In the leftmost panel of Fig. \ref{fig9} two peaks can be distinguished. One  located around  $E/(\omega_o j) = -1$ where $\langle J_z\rangle /j$ takes  its lowest value, and a second one around  $E/(\omega_o j) = 1$  where  $\langle J_z\rangle /j$ takes its largest value. The ESQPTs  are related to the unstable fixed points of the classical Hamiltonian. The second peak is related to the unstable fixed point which appears for any value of the coupling  where the   saturation of the pseudospin variable sets in, whereas  the first  one corresponds to the fixed point which change from stable ($\gamma<\gamma_c$) to unstable at $\gamma=\gamma_c$.

\begin{figure}
\centering{
\begin{tabular}{c}
\rotatebox{90}{\qquad \qquad\qquad \qquad  \Large{$E/(\omega_o j)$}}\includegraphics[angle=0,width=0.45\textwidth]{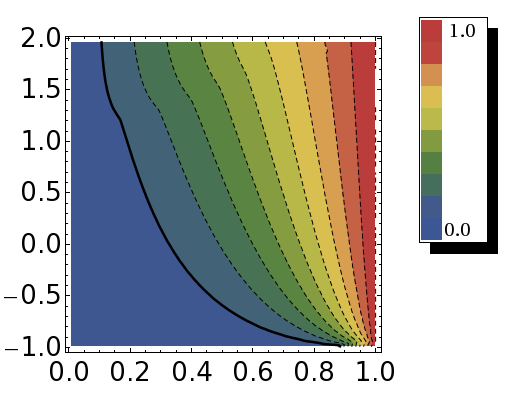}
\\ 
 \quad\rotatebox{90}{\qquad \qquad\qquad   \Large{$E/(\omega_o j)$}}\includegraphics[angle=0,width=0.42\textwidth]{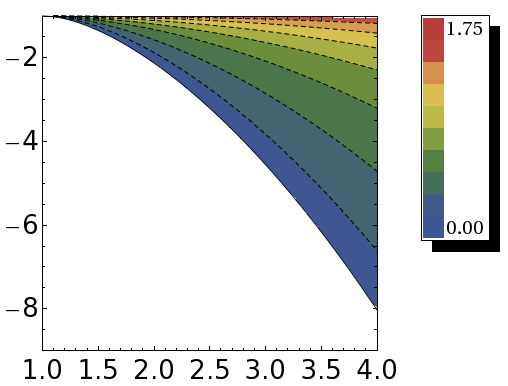}
\\
 \Large{$\gamma/\gamma_c$}
\end{tabular}}
\caption{(Color online) $v_{max}$  [see Eq.(\ref{vv})] as a function of the energy and coupling for  $\gamma< \gamma_c$ (top) and   $\gamma>\gamma_c$ (bottom).}
\label{figVal}
\end{figure}

It is possible to understand the structure of the quantum Peres lattices from the perspective of the classical model. As it is shown in Appendix \ref{app1}, if we make a small oscillations approximation around the energy minimum, a quadratic Hamiltonian is obtained whose normal frequencies are given by   
$$
2 \omega_\pm^2=\omega^2+\omega_o^2\pm \sqrt{(\omega_2^2-\omega_o^2)^2+ 16 \omega\omega_o \gamma^2},$$
for the normal phase and 
$$
2 \gamma_c^4 \omega_{\pm}^2=\omega_o^2\gamma^4+\omega^2\gamma_c^4\pm\sqrt{(\omega_o^2\gamma^4-\omega^2\gamma_c^4)^2+4\omega^2\omega_o^2\gamma_c^8},
$$
for the superradiant one. These   frequencies are equal to those obtained in Ref.\cite{Emary03} by making  a Holstein-Primakhof mapping of the pseudospin variables in the quantum model, likewise they were recently derived by linearizing the classical equation of motion \cite{Bakemeier13}. This latter method is completely equivalent to the one shown in Appendix \ref{app1}. Therefore, for energies close enough to the energy minimum, a two dimensional anisotropic harmonic oscillator is obtained with excitation eigenenergies given by $E_{n_+,n_-}={n_+}\omega_++ n_- \omega_-$, with $n_\pm$ integer numbers equal or greater than zero. The Peres lattice  of such quadratic Hamiltonian will be completely regular, and this is    what it can be seen in the Peres lattices of Fig.\ref{fig9}  for energies close to the ground-state energy. In order to obtain a rough estimate of the range of validity of the quadratic (small oscillations) approximation, we consider for a given energy ($E$) and coupling ($\gamma$), the parameter   
\begin{equation}
v=\left|\frac{E -H_q(q,p,Q_{1},P_{1})}{E-E_{gs}}\right|,
\label{vv}
\end{equation}
where $H_q$ is the quadratic approximation of the semiclassical Hamiltonian  in the normal or superradiant phase (see Appendix \ref{app1}), and 
 $E_{gs}$
 is the  classical ground state energy
 for the given $\gamma$ [see Eq.(15) of companion paper (I)].  
 The smaller is the parameter $v$, the better
 is the quadratic approximation.
 We maximize the previous parameter  in all the available phase space for a given energy $E$ and coupling $\gamma$.  The results are shown in Fig.\ref{figVal} for the normal (top) and superradiant (bottom) phases. 

As it can be seen in the figure,  for very small couplings ($\gamma/\gamma_c\approx 0$) the quadratic approximation is valid for any  energy, but for larger couplings, the range of energy where the quadratic approximation is good  decreases as the coupling approaches to the critical value \cite{Hir13}.  For the critical value the quadratic approximation breaks  completely for any energy because the stable point around which the small oscillation expansion is made changes from stable to unstable (saddle point), and consequently one of the normal modes is equal to  zero.   For couplings above the critical value the  quadratic approximation  is valid only for a small interval above $E_{gs}$, which increases as the coupling does. In the energy intervals where the quadratic approximation is good,  only regular classical orbits  are expected and correspondingly a regular Peres lattice in the quantum version. For energies out of these intervals irregular or chaotic trajectories are expected to emerge.  In order to visualize the way as the regular tori break as a function of energy and coupling, we use Poincare surface sections both for couplings below and above the critical one. These Poincare sections will be compared with the respective results of the quantum model.

\subsection{Poincar\'e Sections and Peres Lattices}

\begin{figure}
\begin{tabular}{c}
\rotatebox{90}{\qquad \qquad\qquad  \Large{$\langle J_z\rangle/j$}}\includegraphics[angle=0,width=0.43\textwidth]{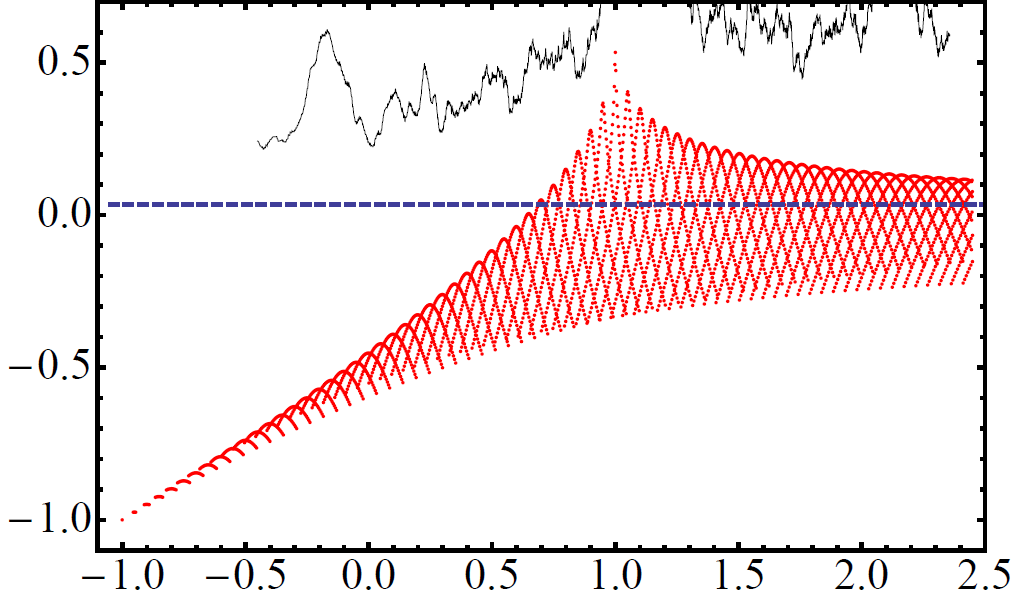}
\\
 \qquad \qquad\Large{$E/(\omega_o j)$}
\end{tabular}
 \caption{(Color online) Peres lattice $j_z/j$ vs $E/(\omega_o j)$  for a finite ($j=40$) Dicke model and  a coupling $\gamma=0.2 \gamma_c$ [a cutoff  $N_{max}=160$ was used, see (I)]. In the same graph, the  Anderson-Darling parameter ($A^2/70$, solid black line)  for a test against the Wigner distribution of the  Nearest Neighbor Spacings  of consecutive 301 states in the spectrum, is shown as a function of the mean energy of the respective states. The horizontal dashed line indicates the maximal value ($2.5/70$) for which the test does not reject the hypothesis of a Wigner distribution for a confidence level of $95 \%$.}
\label{fig12}
\end{figure}
\begin{figure}
\centering{
\begin{tabular}{cc}
\includegraphics[angle=0,width=0.18\textwidth]{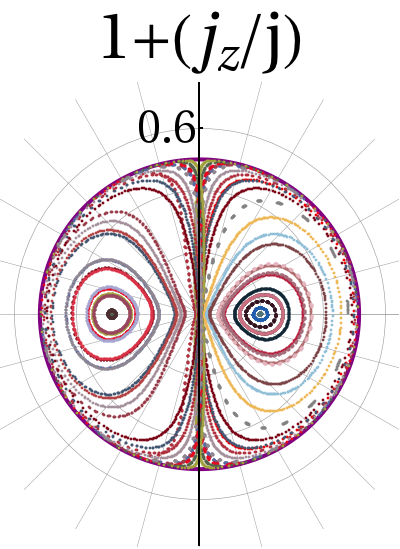}
&\includegraphics[angle=0,width=0.175\textwidth]{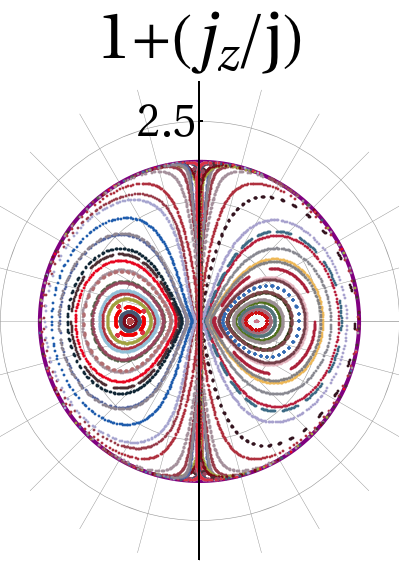}
\end{tabular}}
 \caption{(Color online) Poincar\'e sections ($p=0$)  in the polar  plane  $[1+(j_z/j)]$-$\phi$   of the classical model for the same coupling as previous figure ($\gamma=0.2 \gamma_c$). They correspond to energies  $E/(\omega_o j)=-0.5$ and $2.0$.
 }
 \label{fig12b}
\end{figure}

In this subsection we make a correspondence between the Poincar\'e sections obtained through the semiclassical Hamiltonian, and Peres lattices attained by  numerically diagonalizing  Dicke Hamiltonians with $\omega=\omega_o=1$ and different systems sizes ($j=40$, $j=80$, and $j=100$).  Convergence of the numerical results respect to the cutoff in the bosonic space ($N_{max}$) was checked as it is explained in (I). Only results for the parity positive sector of the model are  presented, but similar results are obtained for the negative parity sector (see Appendix \ref{app2}). Likewise, we analyze  the statistical properties of the quantum energy spectrum in the following way:  for all the cases studied, we consider energy intervals of $N=301$ 
consecutive states with positive parity.
Knowing the density of sates in the classical limit, it is possible to calculate the unfolded ($e_i$) energy spectrum \cite{Haake} as $e_i=\Gamma_+(E_i)$, with $E_i$ the i-th eigenenergy and  $\Gamma_+(E)$ the  cumulative  energy density   given by  $\Gamma_+(E)=\int_{E_{min}}^E \nu_+(E')dE'$.  $\nu_+(E)$ is  the density of parity positive states, which   is given by  $\nu_+(E)=\nu(E)/2$, where $\nu(E)$ is the classical approximation to the density of states  calculated in (I), and given here in  Eq.(\ref{dosDicke}). It is  easy to prove that the  differences of the unfolded energy spectrum can be approximated by   $\Delta_i\equiv e_{i+1}-e_i= (1/2)\nu([E_{i+1}+E_i]/2) (E_{i+1}-E_i)$.  With the unfolded energy differences  for a given interval ($N-1=300$ differences), we test if they follow the Wigner distribution  $P_w(s)=(1/2)\pi s e^{-\pi s^2/4}$,  characteristic of the quantum systems with hard chaotic classical analogue. We use the statistical  Anderson-Darling test \cite{Anderson} which consists of calculating the  so called Anderson-Darling (A-D) parameter 
\begin{eqnarray}
A^2&=&-(N-1)\\
    &-&\sum_{k=1}^{N-1} \frac{2k-1}{N-1}  \left( \ln F_w(\Delta_k)+\ln[ 1-F_w(\Delta_{N-k})]\right),\nonumber
\end{eqnarray}
where the $\Delta_k$ differences are organized  in ascending order,  such that $\Delta_k\leq \Delta_{k+1}$, and $F_w(s)$ is the cumulative distribution function of the Wigner distribution $F_w(s)=\int_{0}^s P_w(s') ds'$. It can be shown \cite{Anderson} that if a set of data ${\Delta_k}$ comes from the theoretical Wigner distribution, the  probability of obtaining a parameter  $A^2$ greater than $2.5$ is 0.05 (Pr$(A^2>2.5)=0.05$). Then if we obtain an A-D parameter larger than 2.5 for a given set of consecutive  $N$ eigenenergies, we can conclude, to a confidence level of 95 \% that the  statistical properties of the energy fluctuations are not described by the Wigner surmise of the quantum chaotic systems.  The A-D parameter can be thought as a measure of the distance of the energy fluctuations of the Dicke energy spectrum  to the Wigner distribution. 

\begin{figure}\begin{tabular}{c}
\rotatebox{90}{\qquad \qquad\qquad  \Large{$\langle J_z\rangle/j$}}\includegraphics[angle=0,width=0.43\textwidth]{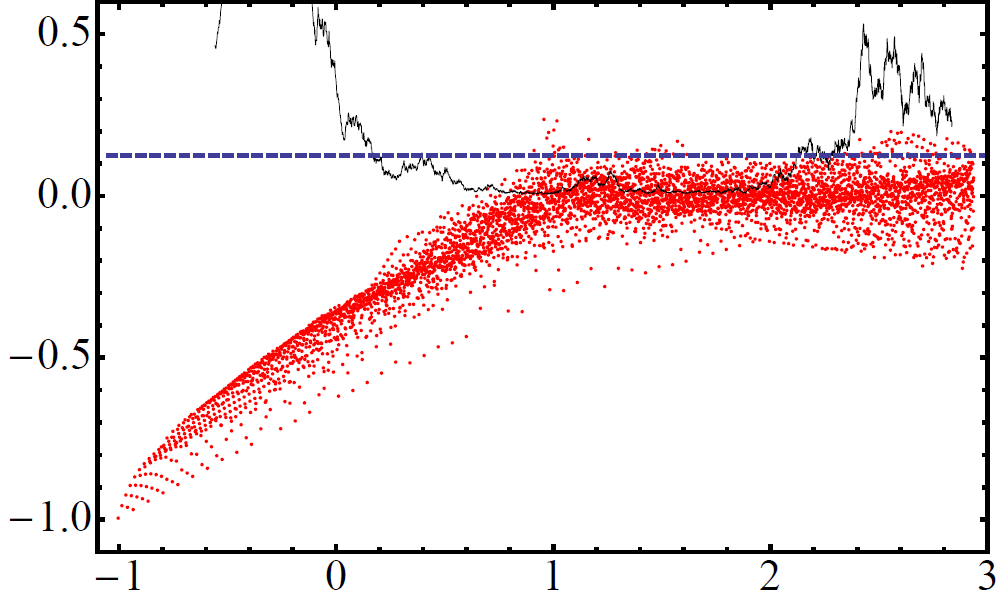} 
\\
 \qquad \qquad\Large{$E/(\omega_o j)$}
\end{tabular}
\caption{(Color online)  The same as  Fig.\ref{fig12},  for a coupling  $\gamma=0.9 \gamma_c$ and an Anderson-Darling parameter divided by 20 ($A^2/20$). The horizontal dashed line indicates the value $2.5/20$. 
}
\label{fig13a}
\end{figure} 
\begin{figure}
\begin{tabular}{cc}
\vtop{\vskip -0.16\textwidth \hbox{\includegraphics[angle=0,width=0.2\textwidth]{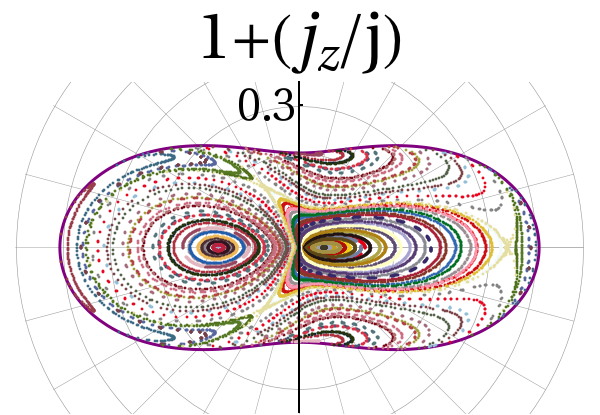}}}
&  \includegraphics[angle=0,width=0.2\textwidth]{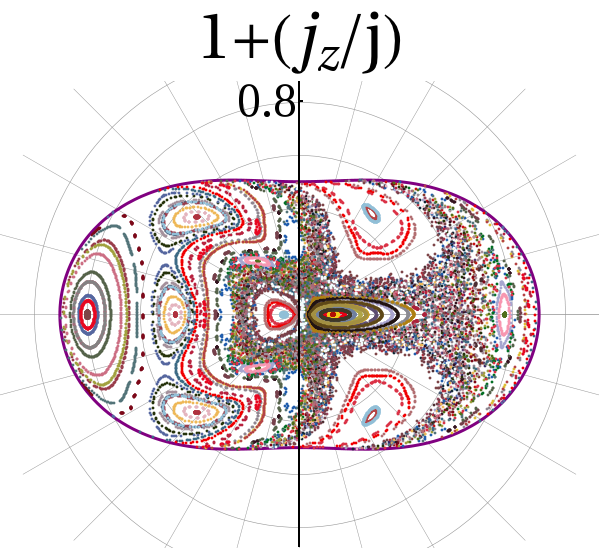}
\\
\vtop{\vskip -0.241\textwidth \hbox{ \includegraphics[angle=0,width=0.2\textwidth]{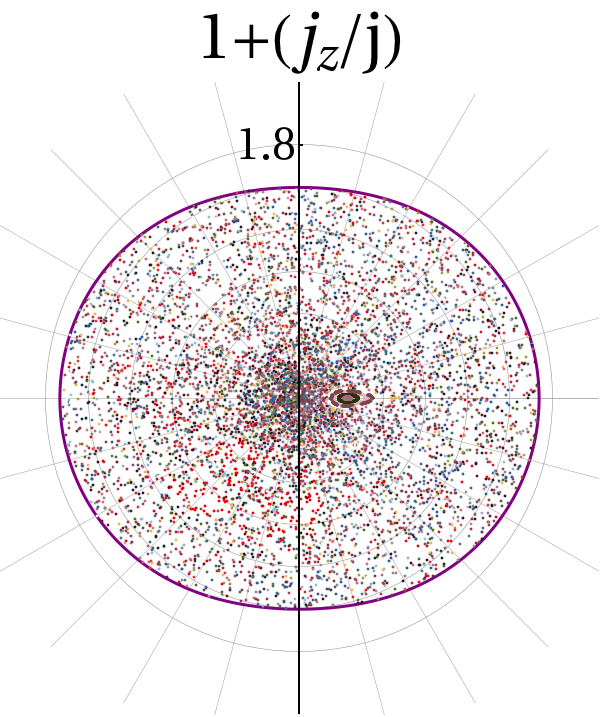}
}}
 & \includegraphics[angle=0,width=0.18\textwidth]{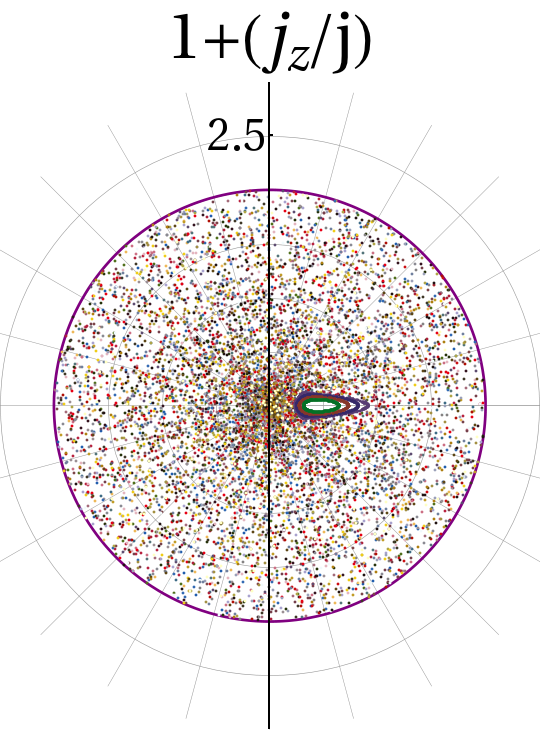}
  \\
 \includegraphics[angle=0,width=0.18\textwidth]{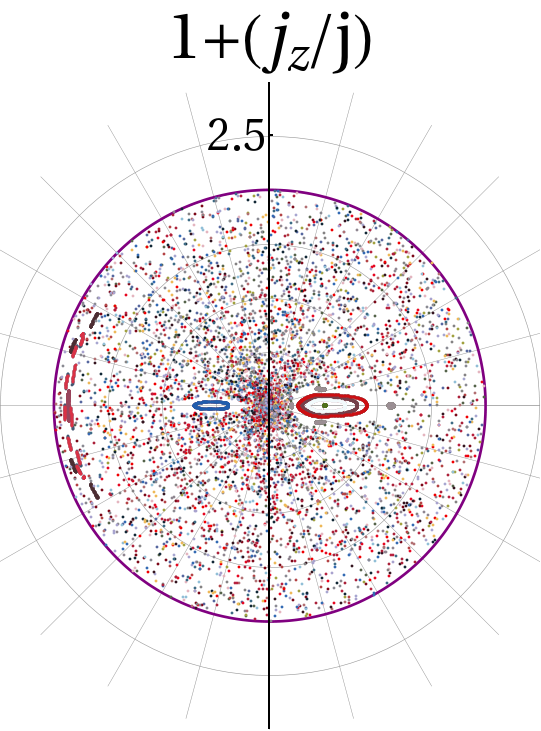}
 & \includegraphics[angle=0,width=0.18\textwidth]{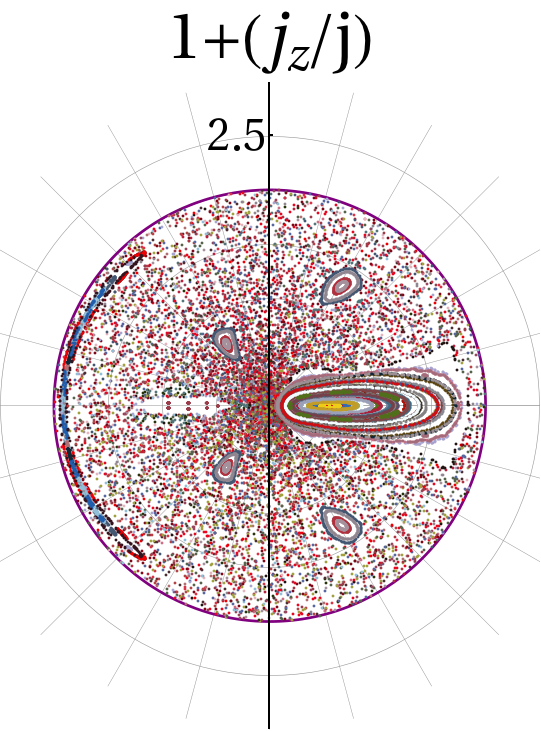}
  \\
\end{tabular}	
\caption{(Color online) Classical Poincar\'e sections ($p=0$)  in the polar  plane  $[1+(j_z/j)]$-$\phi$   for the same system as previous figure ($\gamma=0.9 \gamma_c$) and energies  $E/(\omega_o j)= -0.8$, $-0.5$ (top row), $0.5$, $1.2$ (central row), and  $1.8$, $2.5$ (bottom row).} 
\label{fig13b}
\end{figure}    
We choose representative values of the coupling  in order to observe several regions: the weak coupling normal phase ($\gamma=0.2 \gamma_c$), the normal phase close to the critical value ($\gamma=0.9 \gamma_c$),   the  critical coupling  ($\gamma_c$), the superradiant phase near  the critical coupling ($\gamma=1.35\gamma_c$), and finally, a strong coupling in the superradiant phase ($\gamma=2\gamma_c$)  


 \subsubsection{Normal phase}

In Fig.\ref{fig12}  we present the Peres  lattice $E$ vs $ \langle J_z\rangle/j$  for a small coupling in  the normal phase ($\gamma=0.2 \gamma_c$). As  expected according to the results of  Fig \ref{figVal}, the Peres lattice is completely regular in every energy interval.  This regularity is reflected by  the A-D parameter which is greater than 2.5 for all the energy  intervals.   Correspondingly, the Poincar\'e sections of the classical model (Fig.\ref{fig12b}) show that, independent on energy,  the whole phase space is filled with   regular orbits. The {\it static} ESQPT that takes place at $E/(\omega_o j)=1$, is clearly seen in the Peres lattice as a peak located  at that energy, where the expectation value $\langle J_z \rangle$ attains its maximal value.

Next, we increase the coupling to a value near but below the critical value ($\gamma=0.9\gamma_c$) where, according to Fig.\ref{figVal}, the two modes quadratic approximation is valid only in a small interval above the energy minimum.  In Fig.\ref{fig13a} the corresponding Peres lattice is shown, where it can be observed that only in the low part of the energy spectrum a regular lattice appears, a regularity which is explained by the two modes quadratic approximation around the $E_{gs}$. For larger energies the lattice is completely irregular. The A-D parameter quantifies this change observed in the Peres lattice: for energies close to the ground states it is greater than 2.5 (rejecting, then, the hypothesis of a Wigner distribution in the energy fluctuations), and as the energy increases the A-D parameter decreases. For an energy close to $E/(\omega_o j)=0.2$ it attains  values below $2.5$. 

The corresponding Poincar\'e sections (Fig.\ref{fig13b}) follow closely the previous route of the quantum model. For energies close to the energy minimum, the phase space is covered only by regular orbits. As the energy increases, some regular tori break and a mixed phase space is obtained with regular and chaotic orbits. For energies  $E/(\omega_o j)>0$ the regular tori have  almost disappeared and the phase space is ergodically covered by chaotic trajectories. Interestingly, for energies around $E/(\omega_o j)\approx 2.5$, a  revival of regular orbits is obtained in the classical model (bottom right panel of Fig.\ref{fig13b}), in correspondence this revival is also seen in the quantum model in the same energy region, where a regularity can be observed in the Peres Lattice, and  the A-D parameter increases above $2.5$. The {\it static} ESQPT that takes place at energy $E/(\omega_o j)=1$ can be seen in the Peres lattice even if it is blurred by the quantum chaos present in this energy region. Another interesting  characteristic of the Peres lattice can be seen in energies above $E/(\omega_o j)>1$. According to the classical analysis, for these energies  the trajectories are chaotic  and explore ergodically the whole pseudo-spin sphere. The quantum consequence of this classical result is that the expectation value of the operator $J_z$ (in fact any component of the pseudospin operator $\vec{J}$) must be  equal to zero. This is what can be seen in the Peres  lattice for $E/(\omega_o j)>1$,  where the points are 
noticeably localized
around the value $\langle J_z\rangle=0$.          

\begin{figure}
\begin{tabular}{c}
\rotatebox{90}{\qquad \qquad\qquad  \Large{$\langle J_z\rangle/j$}}\includegraphics[angle=0,width=0.43\textwidth]{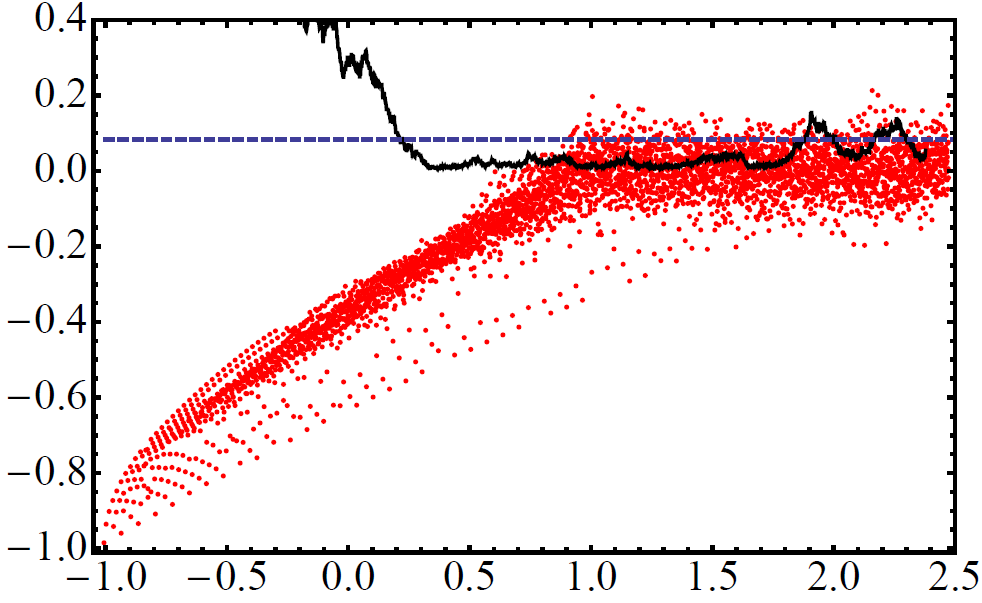} 
\\
 \qquad \qquad\Large{$E/(\omega_o j)$}
\end{tabular}
\caption{(Color online) The same as  Fig.\ref{fig12},  for the critical coupling  $\gamma=\gamma_c$ and an Anderson-Darling parameter divided by 30 ($A^2/30$). The horizontal dashed line indicates the value $2.5/30$. }
\label{critA}
\end{figure}

\begin{figure}
\begin{tabular}{cc}
\includegraphics[angle=0,width=0.2\textwidth]{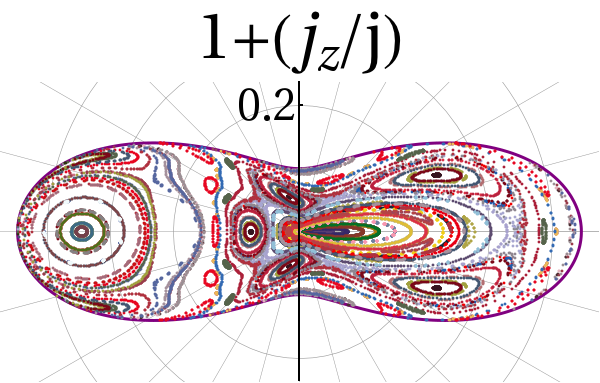}
&\vtop{\vskip -0.125\textwidth \hbox{\includegraphics[angle=0,width=0.21\textwidth]{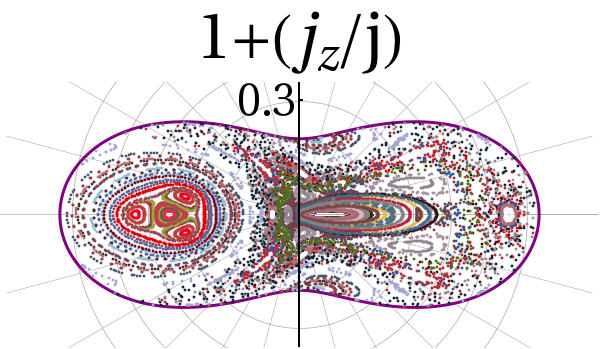}
}}\\
\vtop{\vskip -0.215\textwidth \hbox{\includegraphics[angle=0,width=0.22\textwidth]{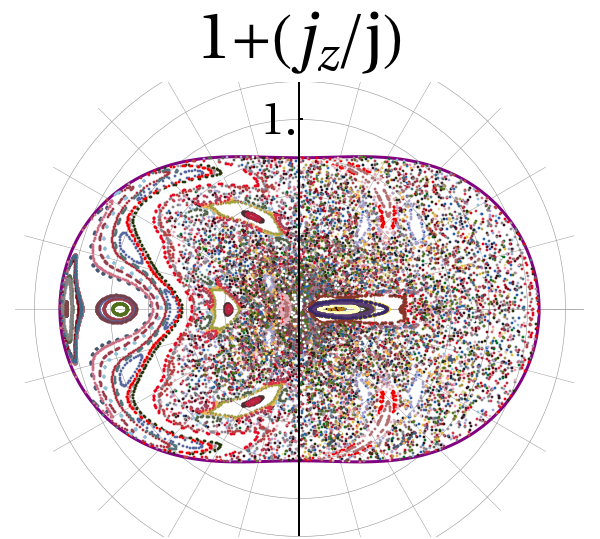}
}}
& \includegraphics[angle=0,width=0.22\textwidth]{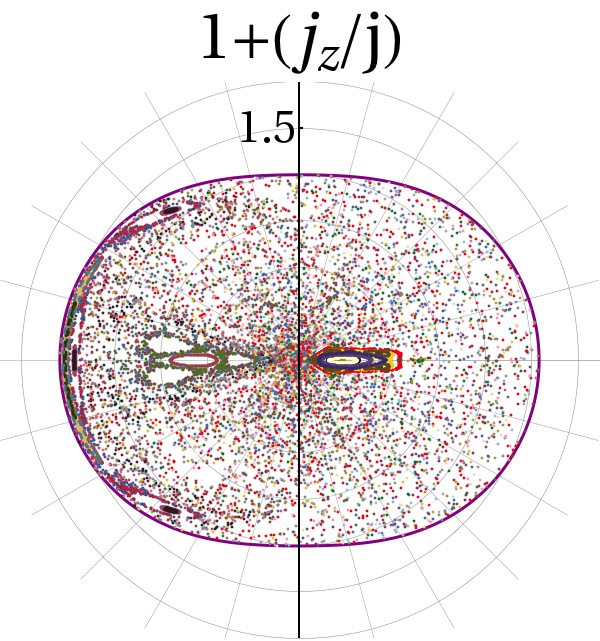}
 \end{tabular}
 \caption{(Color online) Poincar\'e sections ($p=0$) for $\gamma=\gamma_c$ and energies $E/(\omega_o j)=-0.9$, $-0.8$ (top), and  $-0.2$,  $0.2$ (bottom).}
\label{critB}
\end{figure}

The small oscillations approximation allows to explain the regularity observed in the low energy region both in the quantum and classical results, however the breaking of this quadratic approximation does not mean that the system is chaotic. This statement can be clearly seen in the case of the model with the thermodynamical limit critical coupling $\gamma=\gamma_c$. For this particular case the quadratic approximation fails for any energy because  one of its  normal modes  is  exactly zero. Even so, the Peres lattices and the classical trajectories   show (Figs.\ref{critA} and \ref{critB}) regular patterns for energies close to the minimum, with  irregular features appearing at larger energies ($E/(\omega_o j)\gtrsim -0.8$). The results for the critical value  are very similar to those of the case $\gamma=0.9\gamma_c$ discussed above. 

The  above
results  show that 
the presence of chaos in the Dicke model is not restricted to the superradiant phase.
 Irregular patterns, in the classical and corresponding quantum model, appear, except in the perturbative region $\gamma\approx 0$,
in the normal phase for large enough energies.
The same can be said for the critical case, regularity is observed in the energy regime immediately above the energy minimum, and chaotic features are observed at larger energies.    

\subsubsection{Superradiant phase}
\begin{figure}
\centering{
\begin{tabular}{c}
\ \rotatebox{90}{\qquad \qquad\qquad  \Large{$\langle J_z\rangle/j$}}\includegraphics[angle=0,width=0.42\textwidth]{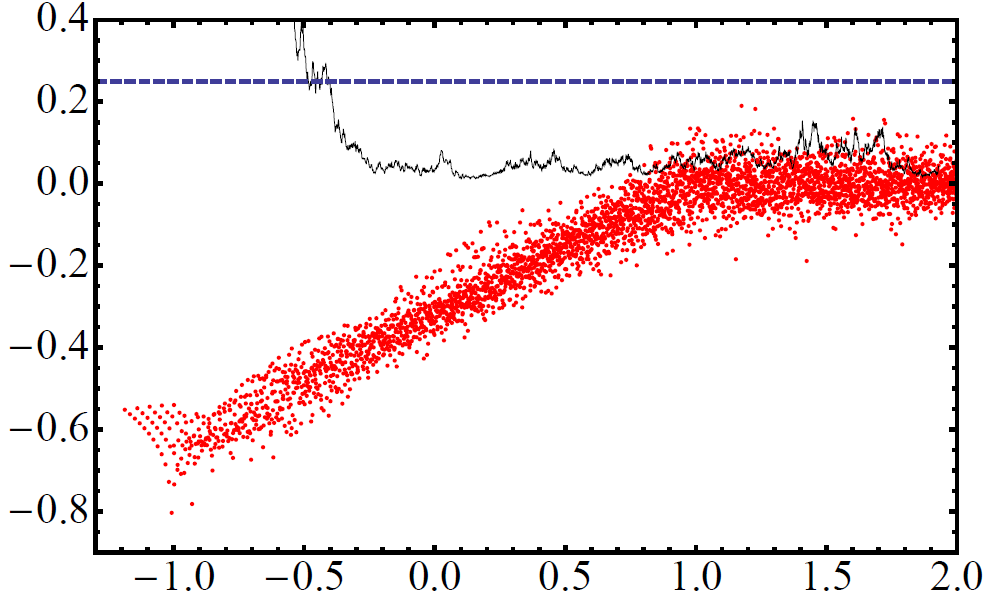}\\
\ \rotatebox{90}{\qquad \qquad\qquad  \Large{$\langle J_z\rangle/j$}}\includegraphics[angle=0,width=0.433\textwidth]{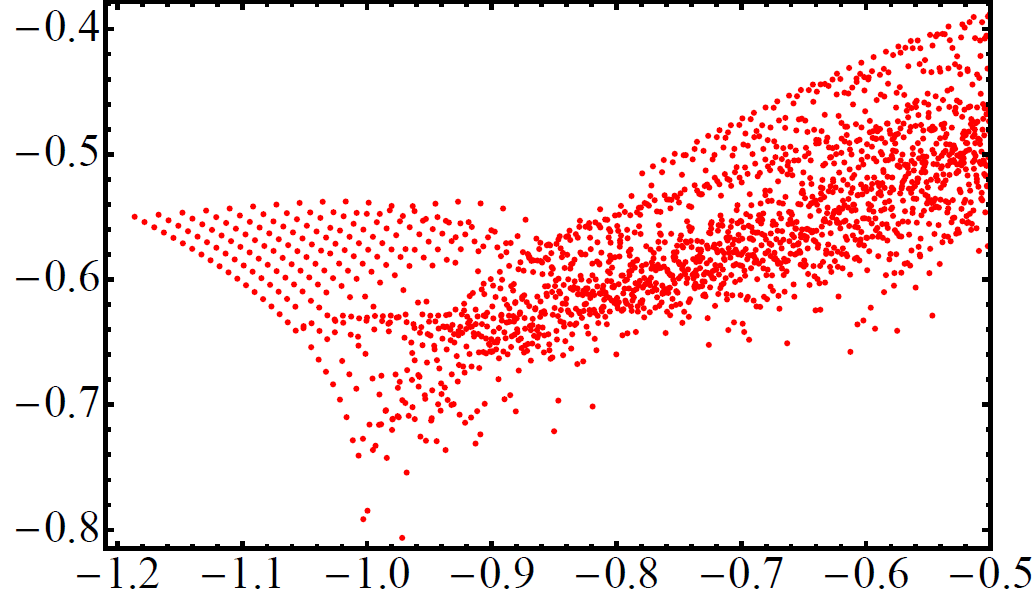}\\
 \quad\rotatebox{90}{\qquad \qquad\qquad  \Large{$A^2$}} \includegraphics[angle=0,width=0.408\textwidth]{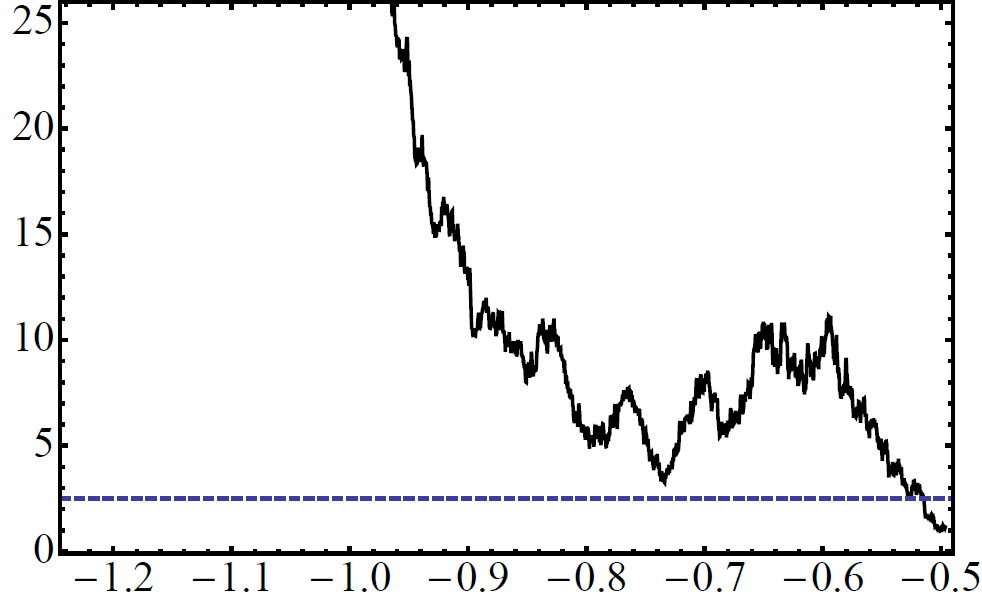}\\
\Large{$E/(\omega_o j)$}
\end{tabular}}
\caption{(Color online)  Top: Peres lattice for  a finite system $j=40$ and coupling $\gamma/\gamma_c=1.35$ (a cutoff $N_{max}=160$ was used) and  the  Anderson-Darling parameter ($A^2/160$, solid black line). The horizontal dashed line indicates the maximal value ($2.5/160$) for which the test does not reject the hypothesis of a Wigner distribution for a confidence level of $95 \%$.
Central: A closer view of the low part of the spectrum for a larger system ($j=100$, with a cutoff $n_{max}=70$). Bottom: Anderson-Darling parameter for the system of central panel, the horizontal dashed line indicates the value $2.5$.}
\label{fig14}
\end{figure}
\begin{figure}
\centering{
\begin{tabular}{cc}
\vtop{\vskip -0.138\textwidth \hbox{ \includegraphics[angle=0,width=0.2\textwidth]{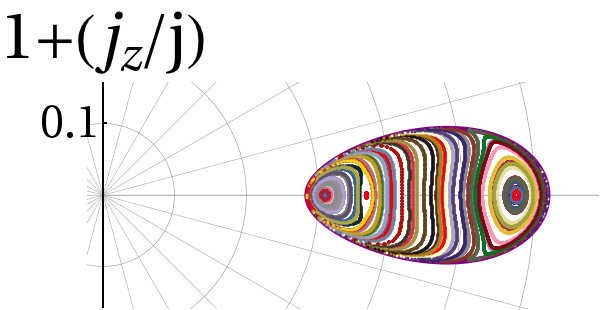}
}}
 & 
 \includegraphics[angle=0,width=0.2\textwidth]{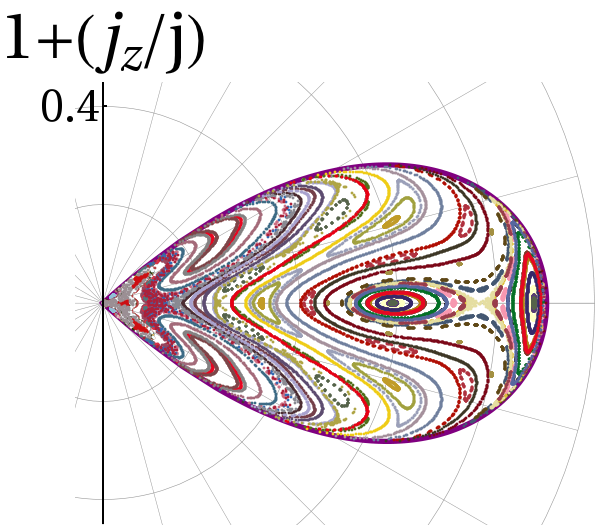}
 \\
 \vtop{\vskip -0.147\textwidth \hbox{\includegraphics[angle=0,width=0.23\textwidth]{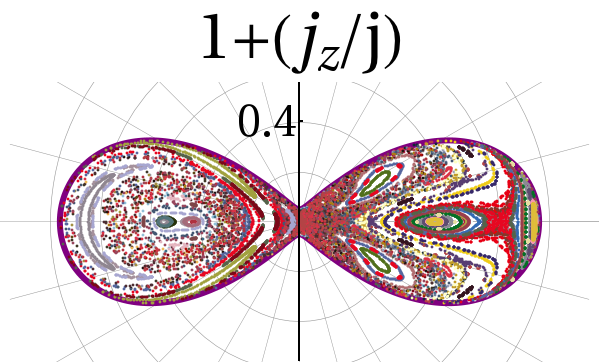}
 }}
 & 
 \includegraphics[angle=0,width=0.23\textwidth]{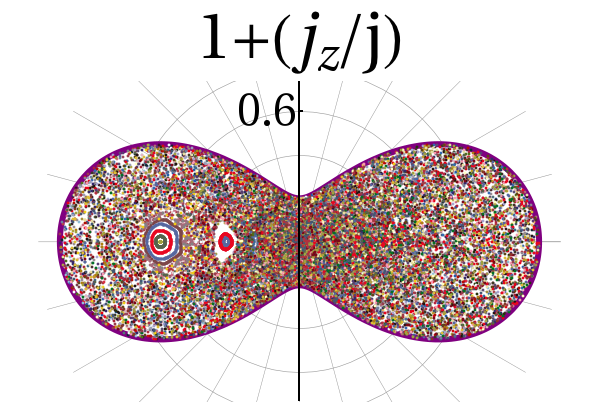}
  \\
\vtop{\vskip -0.16\textwidth \hbox{ \includegraphics[angle=0,width=0.2\textwidth]{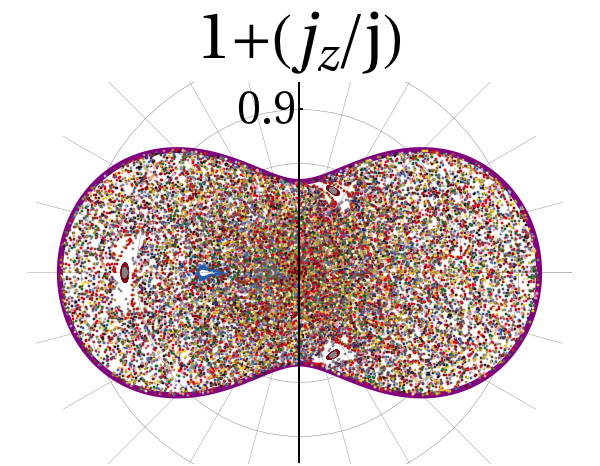}
}}
 & 
 \includegraphics[angle=0,width=0.14\textwidth]{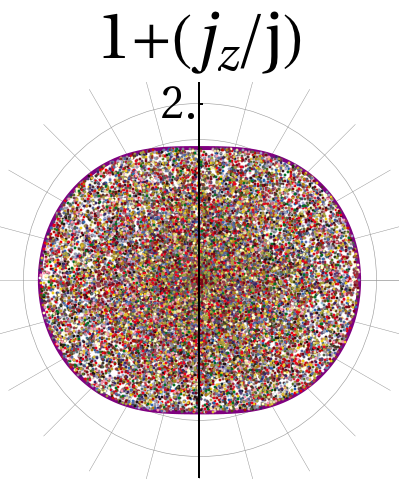}
\end{tabular}}
\caption{(Color online)  Poincar\'e surface sections for the same system as the previous figure ($\gamma=1.35\gamma_c$). Every panel correspond respectively to energies   $E/(\omega_o j)=-1.16$, $-1.0$ (top); $-0.95$,  $-0.8$ (central); and   $-0.5$, $0.5$ (bottom).}
\label{fig15}
\end{figure}

For couplings in the superradiant phase, a new energy region appears below the normal phase lowest energy. This new energy region corresponds, from a classical point of view, to the motion of the pseudospin variables in the double well energy surface where the parity symmetry is spontaneously broken. The top of this double well energy surface is given by the ground state energy of the normal phase ($E/(\omega_o j)=-1$), where the {\it dynamical} ESQPT takes place. As a first example in this coupling regime, we consider a coupling near but above the critical one  $\gamma=1.35\gamma_c$. The corresponding Peres lattices are shown in Fig.\ref{fig14}, whereas the classical Poincar\'e sections are shown in Fig.\ref{fig15}.
As it happened in the previous cases and as expected according to the Fig.\ref{figVal}, for the low energy regime a regular lattice is obtained and correspondingly only classical regular orbits appear in the Poincar\'e sections. In the case of a medium size quantum system ($j=40$ top panel of Fig.\ref{fig14}), the regular  lattice seems to extend until the energy of the {\em dynamic} ESQPT ($E/(\omega_o j)=-1$), however a closer view of the same energy region using a larger system ($j=100$)  unveils a richer structure in the denser Peres lattice: the regular lattice extends beyond the critical energy $E/(\omega_o j)=-1$. For a small interval around  $E/(\omega_o j)=-1$, regular and irregular  lattices coexist. For larger energies the lattice is completely irregular. These lattice characteristics are reflected by the respective  A-D parameter (bottom panel  of Fig.\ref{fig14}) which  decrease drastically from large values at energies close to the minimum.  At  $E/(\omega_o j)\approx -0.8$  it attains values around $A^2=5$, and finally at  an energy $E/(\omega_o j)\approx -0.5$ it takes  values smaller than $A^2=2.5$. The classical results have a clear correspondence with the quantum ones, as can be  verified  in the Poincar\'e sections of  Fig. \ref{fig15}: regular orbits for low energies (including the critical energy $E/(\omega_o j)=-1$), mixed dynamics or soft chaos (coexistence of regular and chaotic trajectories) at energies near but  above  the critical energy $E/(\omega_o j)=-1$, and hard chaos (chaotic trajectories fulling the whole available phase space) for energies $E/(\omega_o j)\gtrsim -0.5$. The precursors of the ESQPTs, both the {\em dynamic} and {\em static},  can be clearly seen in the Peres lattice of the $j=40$ system (top panel of Fig.\ref{fig14}), as a change in the slope of the tendency of the $\langle J_z\rangle$ values as the energy is increased. 

\begin{figure}
\centering{
\begin{tabular}{c}
\  \rotatebox{90}{\qquad \qquad\qquad  \Large{$\langle J_z\rangle/j$}}\ \includegraphics[angle=0,width=0.408\textwidth]{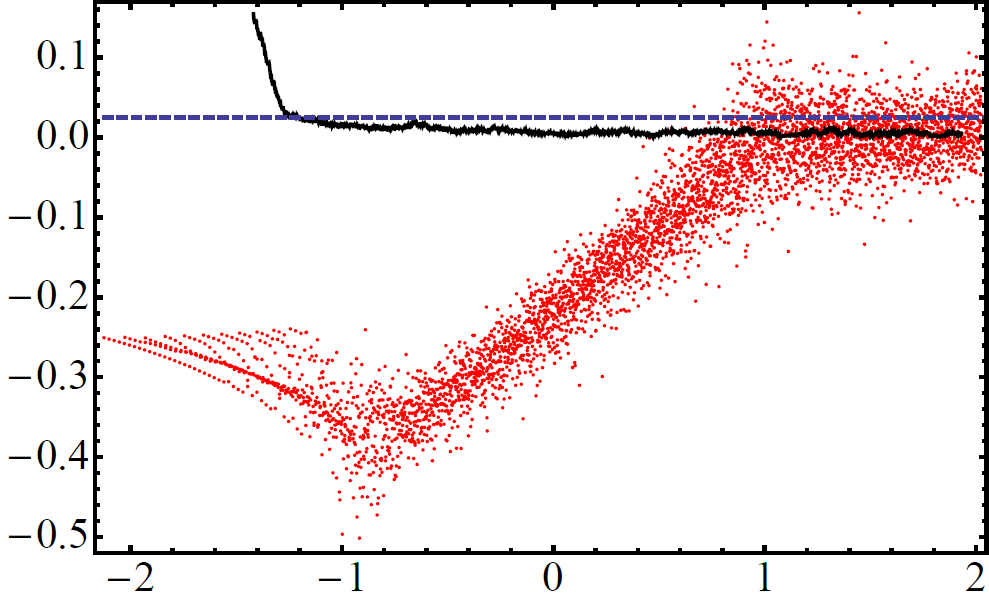}\\
\ \rotatebox{90}{\qquad \qquad\qquad  \Large{$\langle J_z\rangle/j$}}\includegraphics[angle=0,width=0.43\textwidth]{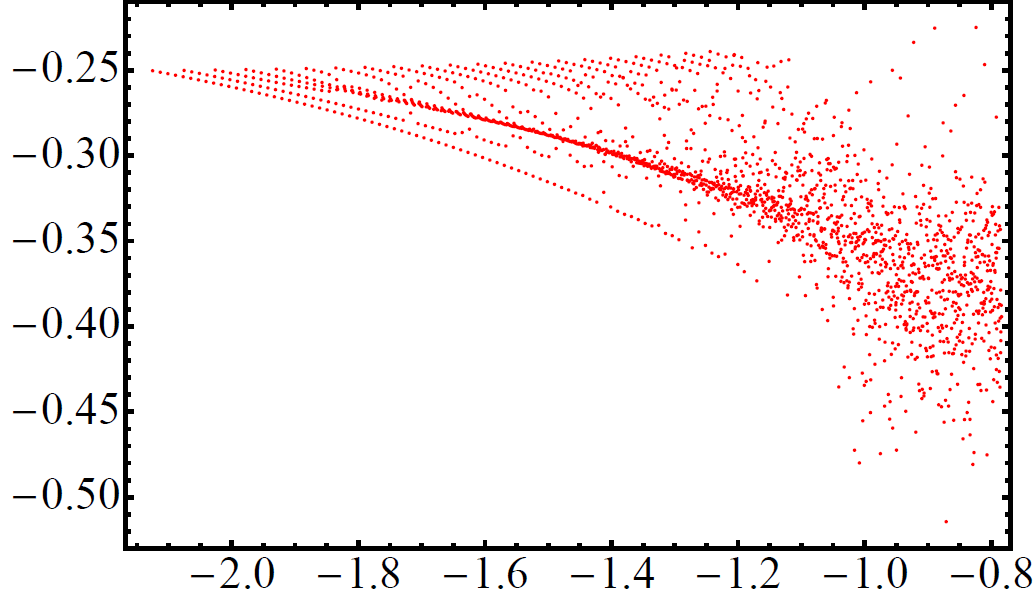}\\
 \quad\quad\rotatebox{90}{\qquad \qquad\qquad  \Large{$A^2$}} \includegraphics[angle=0,width=0.4\textwidth]{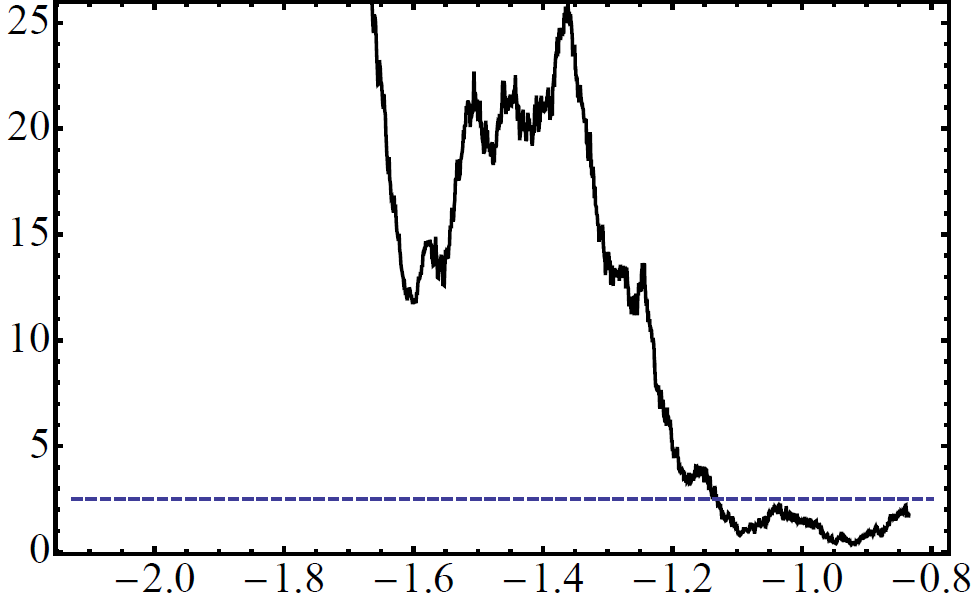}\\
\Large{$E/(\omega_o j)$}
\end{tabular}
}\caption{(Color online) Top: The same as top panel of  Fig.\ref{fig14} but for a coupling $\gamma/\gamma_c=2.0$  and  an   Anderson-Darling parameter divided by 100 ($A^2/100$). The horizontal dashed  line indicates the value $2.5/100$. Central:  Peres Lattice in a smaller energy interval for a larger system ($j=80$), the bosonic cutoff used was  $N_{max}=95$. Bottom: Anderson-Darling parameter for the system of central panel,  the horizontal dashed line indicates the value $2.5$.}
\label{fig16}
\end{figure}

\begin{figure}
\centering{
\begin{tabular}{cc}
 \includegraphics[angle=0,width=0.22\textwidth]{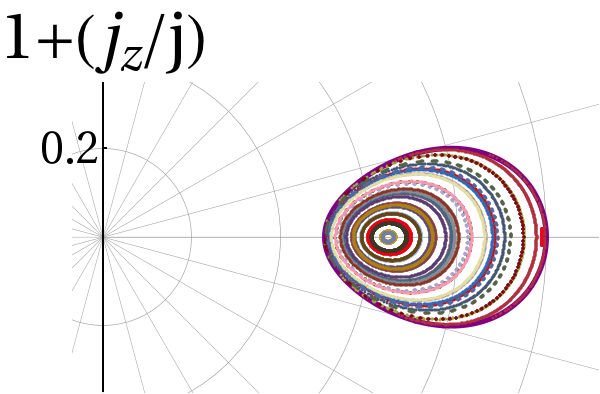}
 & 
 \includegraphics[angle=0,width=0.20\textwidth]{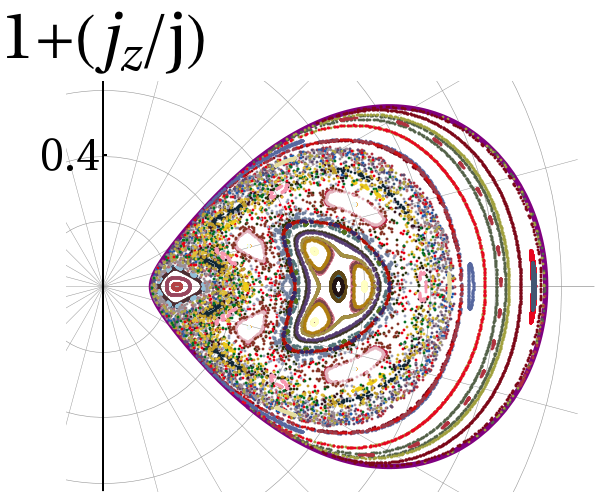}
 \\
 \includegraphics[angle=0,width=0.20\textwidth]{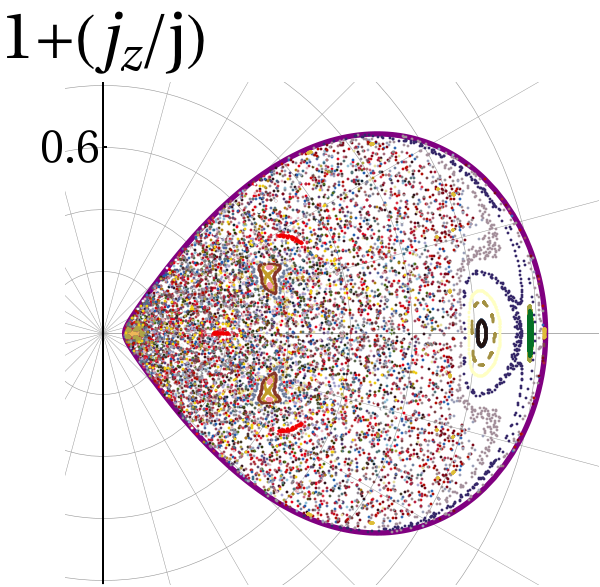}
 &  
 \includegraphics[angle=0,width=0.20\textwidth]{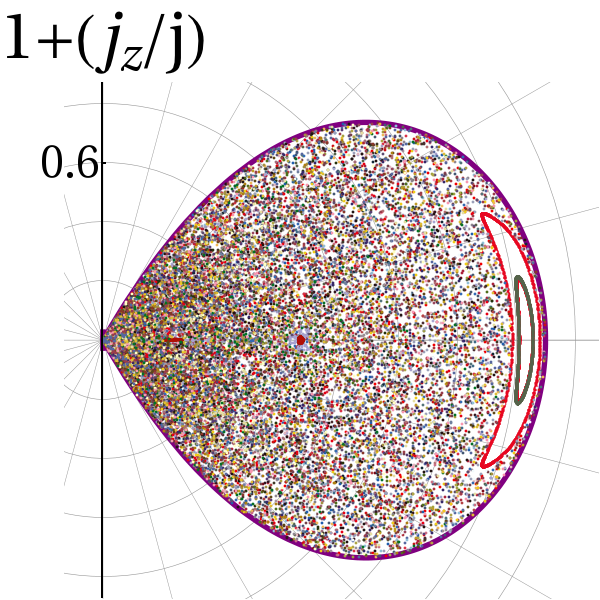}
  \\
 \vtop{\vskip -0.177\textwidth \hbox{\includegraphics[angle=0,width=0.2\textwidth]{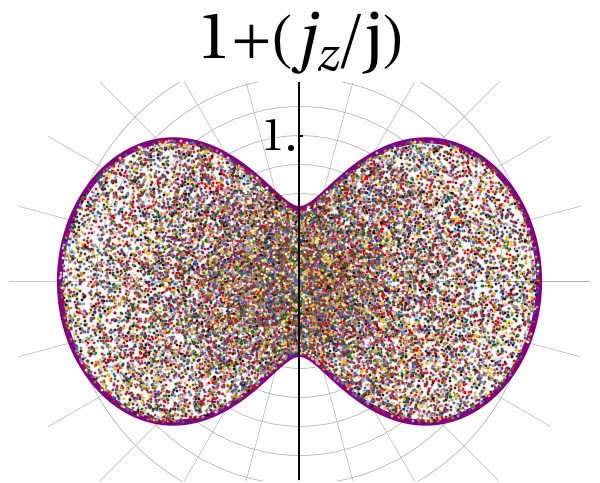}
 }}
  & 
 \includegraphics[angle=0,width=0.14\textwidth]{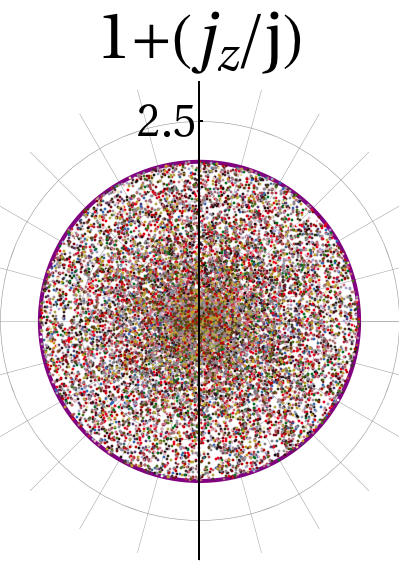}
 \end{tabular}}
 \caption{(Color online)  Poincar\'e surface sections for the same system as the previous figure ($\gamma/\gamma_c=2.0$). Every panel correspond respectively to energies   $E/(\omega_o j)=-2.0$,$-1.4$ (top),  $-1.2$, $-1.0$ (central), and $-0.5$, $1.5$ (bottom). }
 \label{fig17}
 \end{figure}

The last case we present is one with a large coupling,  deep in the superradiant phase  ($\gamma=2.0\gamma_c$). Again, we have  a regular region in the low energy sector and a hard chaotic region for higher energies. As in the previous cases the precursors of the two ESQPTs are clearly seen in the Peres lattice shown in the top panel of Fig.\ref{fig16}. In the same panel  the regular part of the lattice seems to be limited by the critical energy of the {\it dynamic} ESQPT ($E/(\omega_o j)=-1$),  however a closer view to that energy region for a larger system ($j=80$) unveils a more involved relationship between the ESQPT and the transition to chaos. Differently to the case  shown in the central panel of Fig.\ref{fig14} ($\gamma=1.35 \gamma_c$,  for a  system size $j=100$), the regular part of the Peres lattice in this case  (central panel of Fig.\ref{fig16}) does not extend beyond $E/(\omega_o j)=-1$, but it is upper limited by $E/(\omega_o j)\approx -1.1$. For low energies the  lattice is completely regular, whereas  for energies around $E/(\omega_o j)\approx -1.1$ a coexistence of regular and irregular patterns is obtained. At  the critical energy  of the {\em dynamic} ESQPT    the lattice seems to be completely irregular, and the same is obtained for larger energies.  The previous qualitative observations are quantitatively  reflected by the A-D parameter  (bottom  panel of Fig.\ref{fig16})  which decrease abruptly in an  energy interval around  $E/(\omega_o j)\approx -1.1$ below the value $A^2=2.5$. The classical Poincar\'e sections (Fig. \ref{fig17}) present a very similar route to chaos. For $ E/(\omega_o j)=-2.0$ only regular orbits are obtained, whereas for  $E/(\omega_o j)=-1.4$ a mixed phase space with regular and chaotic trajectories appears.  In a narrow energy interval the phase space properties change  rapidly. For  $E/(\omega_o j)=-1.2$ the chaotic trajectories fill  the available  phase space, excepting a small stability island, which  disappears completely at larger energies.

The Peres lattices and the classical results which support and explain the different structures found in them, show that the change from a regular regime to a chaotic one in the Dicke model is much more involved that the transitions linked to the DoS (ESQPT) or the ground-state properties (QPT). While the ESQPT and QPT are well defined transitions in the classical limit [they are unambiguously indicated by non-analytic behavior of the ground-state energy and volume of the available phase space $\nu(E)$], the transition from a regular regime (at low energies) to a chaotic one (large energies) can not be unambiguously defined even in the classical limit. Instead   continuous changes in the phase space structures are obtained, which include a mixed or soft chaos regime  where 
 regular and chaotic patterns coexist,
both in the classical and quantum results, and both in the normal and  superradiant phases. Nevertheless,  an indirect connection between chaos and the ESQPT can be obtained through the unstable fixed points of the classical version.  They determine clearly the energies where changes in the available phase space take place, and consequently singular behavior in the classical approximation of the DoS, on the other hand they are also involved in the breaking of the quadratic approximation which is a necessary (but not sufficient) condition for the presence of chaos in the classical model.             

Finally,  it is worth mentioning that the results presented above show that the Peres lattices are a very useful tool to explore in a qualitative and quick way the presence of chaos in the quantum models. Moreover, they  are useful to give indications about the properties of the corresponding classical model. The Peres lattices allow at a glance to visualize the system properties in the energy space, contrary to the traditional Poincar\'e sections  which are defined  only for a given energy.  With the Peres lattices  it is easy to select individual states and identify if they are part of the regular or chaotic lattice part. In this sense  they can be useful to perform more detailed studies about the classical and quantum correspondence such as the work of Ref.\cite{Bakemeier13}, where the Husimi functions of selected eigenstates are calculated and compared with the classical results. As in the present study, in that Ref.  a clear quantum-classical correspondence is obtained for systems with sizes of order 
$\mathcal{N} \sim \mathcal{O}(10^2)$. 


\section{Conclusions}

Using both a semiclassical analysis and  results of an efficient numerical procedure   to diagonalize the quantum Hamiltonians,  we have studied  the Dicke and TC models in the space of coupling and energy. We focused on the onset of chaos in the non-integrable Dicke model, one of the global properties of the energy-coupling space.

  We explore the properties of the quantum and classical models as a function of coupling and energy with regard to the onset of irregular patterns. Poincar\'e surface sections and Peres lattices were used, respectively,  for the classical and quantum versions. A clear classical and quantum global  correspondence was obtained for system sizes ranging from $\mathcal{N}=80$ to $200$. 
  
  Through the unstable fixed points, an indirect connection between the ESQPT's and the onset of chaos was identified, however, the latter is 
  a much richer phenomenon than the occurrence of non-analytic behavior in the density of states or the ground-state energy. It was found that the onset of chaos is related with the breaking of the quadratic approximation of the Hamiltonian that is obtained by considering small oscillations around the global energy minimum. It was confirmed  in the quantum and classical versions, that chaos is present, both  in the normal and  superradiant phase,  for large enough energies, except in the perturbative regime $\gamma\approx 0$.  Conversely, for any coupling there always exist an  energy interval above the energy minimum where  only regular patterns are obtained. In particular for the very small coupling regime $\gamma\approx 0$ this energy interval extends to infinity.  Once the quadratic approximation is broken more energy is needed to produce chaotic patterns, something that can be clearly seen in the classical system with critical coupling, where one of the normal modes of the quadratic approximation is exactly zero, which implies that terms of order larger than two have to be considered. Even so, the system presents regular patterns in  the low energy regime, and irregular trajectories  appears until larger energies.  The Peres lattices  used to study the quantum versions were a very useful tool to identify qualitatively the chaotic and regular features of the spectrum, moreover they show  clear signatures of the ESQPTs.  The qualitative information provided by the Peres lattices was quantitatively confirmed by analyzing the statistical properties of the quantum fluctuations. It was tested if the fluctuations of  spectrum  in different energy intervals follow the Wigner distribution characteristic of the hard chaotic systems. 
  
The classical analysis performed in this contribution allows us to gain many insights about the results obtained in the quantum versions. For instance, it was shown that the two modes approximation obtained in \cite{Emary03} by making a Holstein-Primakhof of the pseudospin variables in the quantum model, is valid in the thermodynamic (equivalent here to the semi-classical) limit, but only for an energy region immediately above the ground-state energy. Moreover, the two modes approximation explains very well the regular patterns found in the  low lying energy spectrum of the  finite quantum Dicke model.  

  This global study may be useful as a navigation chart to more detailed studies that focus on the classical-quantum correspondence of single states, such as that performed in Ref.\cite{Bakemeier13}. Finally, the results presented here for optical models confirm  results of previous studies,  performed in the context of nuclear  physics simple models \cite{Str09,CejStra1,CejStra2},  about  the classical and quantum correspondence with regard to  the onset of  chaos in the extended energy and coupling space.

We thank P. Str\'ansky and P. Cejnar for many useful and interesting conversations.This work was partially supported by CONACyT- M\'exico,  DGAPA-UNAM and DGDAEIA-UV through the "2013 Internal call for  strengthening  academic groups" (UV-CA-320).


\appendix

\section{Small oscillations around the energy minima}
\label{app1}
If  small oscillations around the energy minimum are considered, a quadratic Hamiltonian is obtained  whose normal frequencies $\omega_\pm$ give the low lying energy spectrum of the quantum model with excitation energies   $E_{n_+,n_-}=\omega_+ n_+ +\omega_-n_- $, with $n_\pm=0,1,2,..$.  For couplings below the critical value the normal modes of the low energy regime can be   obtained  by expanding the classical Hamiltonian (\ref{DiHam}) around the global minimum $j_z=-j$. This expansion is easily obtained by  transforming  \cite{DickeBrasil}  the angular momentum canonical variables ($j_z$ and $\phi$) to $Q_1=\sqrt{2 (j+j_z)}\sin\phi$, $P_1=\sqrt{2 (j+j_z)}\cos\phi$. In terms of these variables  the  classical Hamiltonian (\ref{DiHam}) of  the Dicke model ($\delta=1$),  reads 
\begin{eqnarray} 
H_{cl}&=&-\omega_o j+\frac{\omega_o}{2}(Q_1^ 2+P_1^2)+\frac{\omega}{2}(q^ 2+p^2)\nonumber\\
& &+2\gamma q P_1\sqrt{1-\frac{Q_1^2+P_1^2}{4j}}.
\label{hamQP}
\end{eqnarray}    
By expanding the square root in the previous Hamiltonian, we obtain, to leading order,    a quadratic Hamiltonian 
\begin{equation} H_q=-\omega_o j+\frac{\omega_o}{2}(Q_1^ 2+P_1^2)+\frac{\omega}{2}(q^ 2+p^2)+2\gamma q P_1,
\end{equation}
with normal frequencies  \cite{Goldstein} given by $$2 \omega_\pm^2=\omega^2+\omega_o^2\pm \sqrt{(\omega_2^2-\omega_o^2)^2+ 16 \omega\omega_o \gamma^2}.$$

For couplings larger than the critical one, two degenerate minima emerge, and the expansion has to be taken around these new minima ($q_m$, $p_m$, $\phi_m$, $j_{zm}$)  given by  Eq.(12) in the companion paper (I).  The expansion until the quadratic leading terms is
\begin{eqnarray}
H_q &=& E_{gs}+\frac{\omega}{2}\left[ (q-q_m)^2+p^2\right]\\
 &+&\frac{j\omega_o}{2} \left[ \left( \frac{\gamma^2}{\gamma_c^2}-\frac{\gamma_c^2}{\gamma^2}\right)\phi^2+ \frac{ (\gamma/\gamma_c)^4}{\left( \frac{\gamma^2}{\gamma_c^2}-\frac{\gamma_c^2}{\gamma^2}\right)} \left(\frac{j_z-j_{zm}}{j}\right)^2\right]\nonumber\\
 &+&\frac{\sqrt{\omega\omega_o j}}{\sqrt{\frac{\gamma^2}{\gamma_c^2}-\frac{\gamma_c^2}{\gamma^2}}} (q-q_m)\left(\frac{j_z-j_{zm}}{j}\right). \nonumber  
\end{eqnarray}   

The normal modes of the previous quadratic Hamiltonian can be obtained easily \cite{Goldstein}, they are
$$
2 \gamma_c^4 \omega_{\pm}^2=\omega_o^2\gamma^4+\omega^2\gamma_c^4\pm\sqrt{(\omega_o^2\gamma^4-\omega^2\gamma_c^4)^2+4\omega^2\omega_o^2\gamma_c^8}.
$$


\section{Basis with defined parity}
\label{app2}
The extended bosonic basis  we used in (I) to diagonalize the Dicke  Hamiltonian is given  \cite{Chen0809,Basta11} by the eigenstates of the  Dicke Hamiltonian in the limit $\omega_{0}\rightarrow 0$, which are
\begin{equation}
|N;j,m'\rangle \equiv \frac{1}{\sqrt{N!}}(A^\dagger)^N |N=0; j,m'\rangle,
\label{cohbas}
\end{equation}
where $A^\dagger =a^\dagger+\frac{2\gamma}{\sqrt{\mathcal{N}}\omega}J_{x}$, $m'$ are the eigenvalues of $J_{x}$, and $|N=0; j,m'\rangle=|\alpha=-\frac{2 \gamma m'}{\omega \sqrt{\mathcal{N}}}\rangle |j m'\rangle$, with $|\alpha\rangle$ a boson coherent state and $|j m'\rangle$ an eigenstate of the $J_x$ operator.  The previous states are not eigenstates of   the parity operator $\Pi=e^{i\pi\Lambda}=e^{i\pi (J_z+j) }e^{i \pi a^{\dagger}a}$. 
 In order to analyze the statistical properties of the Dicke spectrum, we have to separate the energy eigenstates according to their  parity  ($p=\pm$). To this end,  we construct a  basis which  is  also an eigenbasis of the parity operator. 
It is easy to prove  that 
\begin{equation}
\begin{split}
&|N;j,m'\rangle  =\\
&\frac{1}{\sqrt{N!}}\left(a^\dagger+\frac{2\gamma}{\sqrt{\mathcal{N}}\omega}m'\right)^N \left|\alpha=-\frac{2 \gamma m'}{\omega \sqrt{\mathcal{N}}}\right\rangle |j m'\rangle,
\end{split}
\end{equation} 
this result shows that the states (\ref{cohbas}) are proportional to  $|jm'\rangle$. It can be shown  (see \cite{Edmonds}) that   the action of the  rotation operator $e^{i\pi (J_z+j) }$ over $|j m'\rangle$ gives $$e^{i\pi (J_z+j)}|j m'\rangle=|j -m'\rangle.$$ Therefore  we have 
\begin{equation}
\begin{split}
 &e^{i\pi (J_z+j) }|N;j,m'\rangle=\\
 &  \frac{1}{\sqrt{N!}}\left(a^\dagger+\frac{2\gamma}{\sqrt{\mathcal{N}}\omega}m'\right)^N \left|\alpha=-\frac{2 \gamma m'}{\omega \sqrt{\mathcal{N}}}\right\rangle |j -m'\rangle. 
\end{split}  
   \label{acrot}
\end{equation}
On the other hand, by using the properties of the coherent states, it is straightforward to show that    $e^{i\pi a^{\dagger}a}(a^{\dagger})^{k}|\alpha\rangle=(-1)^{k}(a^{\dagger})^{k}|-\alpha\rangle$. With the  previous result we obtain 
\begin{equation}  
\begin{split}
&e^{i\pi a^{\dagger}a}|N;j,m'\rangle=\\
&(-1)^{N}\frac{1}{\sqrt{N!}}\left(a^\dagger-\frac{2\gamma}{\sqrt{\mathcal{N}}\omega}m'\right)^N \left|\alpha=\frac{2 \gamma m'}{\omega \sqrt{\mathcal{N}}}\right\rangle |j m'\rangle.
\end{split}
\label{p3}
\end{equation}
By putting together Eqs.(\ref{acrot}) and (\ref{p3}) we obtain 
\begin{eqnarray}  
\Pi|N;j,m'\rangle&  \nonumber\\
=(-1)^{N}\frac{1}{\sqrt{N!}}\left(a^\dagger-\frac{2\gamma}{\sqrt{\mathcal{N}}\omega}m'\right)^N \left|\alpha=\frac{2 \gamma m'}{\omega \sqrt{\mathcal{N}}}\right\rangle |j -m'\rangle &  \nonumber\\
=(-1)^{N}\frac{1}{\sqrt{N!}}\left(a^\dagger+\frac{2\gamma}{\sqrt{\mathcal{N}}\omega}J_x\right)^N \left|\alpha=\frac{2 \gamma m'}{\omega \sqrt{\mathcal{N}}}\right\rangle |j -m'\rangle &  \nonumber\\
=(-1)^{N} \frac{1}{\sqrt{N!}}(A^\dagger)^N |N=0;j,-m'\rangle= (-1)^{N} |N; j,-m'\rangle .&\nonumber
\label{pf}
\end{eqnarray}
Then, the invariant subspaces of the  the parity operator are generated by  states (\ref{cohbas}) with the same values $N$ and $|m'|$. 
It is straightforward to diagonalize the parity operator in these subspaces, and we obtain the eigenstates of the Dicke Hamiltonian in the limit $\omega_{0}\rightarrow 0$, which are simultaneously eigenstates of the parity operator $\Pi$, 
\begin{equation}
\begin{split}
&|N;j,m';p=\pm \rangle=\\
&=\frac{1}{\sqrt{2(1+\delta_{m',0})}}\left(|N;j,m'\rangle \pm (-1)^{N}|N;j,-m'\rangle\right).
\end{split}
\end{equation}
Using this basis we can  separate from the beginning the two parity sectors of the Dicke model and use the  extended coherent basis, which has been shown \cite{Chen0809,Basta11}  to be very efficient to study large Dicke systems. 



\end{document}